\documentclass[aps,prd,reprint,showpacs, preprintnumbers,amsmath,amssymb,superscriptaddress, nofootinbib]{revtex4-1}
\usepackage{amssymb, amsmath, braket, nicefrac, graphicx} 
\usepackage{relsize}
\usepackage{lineno}
\usepackage{setspace}

\usepackage{esint}
\usepackage{chngpage}
\usepackage{epsfig}
\usepackage{multirow}
\usepackage{rotating}
\usepackage{float}
\usepackage{color}
\usepackage{xcolor}
\usepackage{appendix}
\usepackage{ amstext }
\usepackage{ gensymb }
\usepackage{ textcomp }
\usepackage{ wasysym }
\usepackage{pstricks}

\usepackage[caption=false,labelformat=parens]{subfig}

\usepackage{dcolumn}
\usepackage{bm}
\usepackage{units}

\newcommand{\numu}{\mbox{$\nu_{\mu}$}}                         
\newcommand{\nue}{\mbox{$\nu_{e}$}}                                

\newcommand{\anu}{\ensuremath{\bar{\nu}}}

\newcommand{\anumu}{\ensuremath{\bar{\nu}_{\mu}}}

\newcommand{\piz}{\mbox{$\pi^{0}$}}       
\newcommand{\simgt}{\,\hbox{\lower0.6ex\hbox{$\sim$}\llap{\raise0.6ex\hbox{$>$}}}\,}
\newcommand{\simlt}{\,\hbox{\lower0.6ex\hbox{$\sim$}\llap{\raise0.6ex\hbox{$<$}}}\,}

\definecolor{maroon}{RGB}{162,10,10}

\usepackage{hyperref}
\usepackage[all]{hypcap}
\hypersetup{
    colorlinks,
    citecolor=black,
    filecolor=black,
    linkcolor=black,
    urlcolor=black
}

\makeatletter
\renewenvironment{table}
  {\def\@captype{table}}
  {}

\renewenvironment{figure}
  {\def\@captype{figure}}
  {}
\makeatother
   
\begin{document}
\preprint{FERMILAB-PUB-18-578-ND}
\title{Measurement of $\anumu$ charged-current single $\pi^{-}$ production on hydrocarbon in the few-GeV region using MINERvA}


\newcommand{\Rutgers}{Rutgers, The State University of New Jersey, Piscataway, New Jersey 08854, USA}
\newcommand{\Hampton}{Hampton University, Dept. of Physics, Hampton, VA 23668, USA}
\newcommand{\Dortmund}{Institute of Physics, Dortmund University, 44221, Germany }
\newcommand{\Otterbein}{Department of Physics, Otterbein University, 1 South Grove Street, Westerville, OH, 43081 USA}
\newcommand{\JMU}{James Madison University, Harrisonburg, Virginia 22807, USA}
\newcommand{\Florida}{University of Florida, Department of Physics, Gainesville, FL 32611}
\newcommand{\UCIrvine}{Department of Physics and Astronomy, University of California, Irvine, Irvine, California 92697-4575, USA}
\newcommand{\CBPF}{Centro Brasileiro de Pesquisas F\'{i}sicas, Rua Dr. Xavier Sigaud 150, Urca, Rio de Janeiro, Rio de Janeiro, 22290-180, Brazil}
\newcommand{\PUCP}{Secci\'{o}n F\'{i}sica, Departamento de Ciencias, Pontificia Universidad Cat\'{o}lica del Per\'{u}, Apartado 1761, Lima, Per\'{u}}
\newcommand{\INRM}{Institute for Nuclear Research of the Russian Academy of Sciences, 117312 Moscow, Russia}
\newcommand{\Jlab}{Jefferson Lab, 12000 Jefferson Avenue, Newport News, VA 23606, USA}
\newcommand{\Pittsburgh}{Department of Physics and Astronomy, University of Pittsburgh, Pittsburgh, Pennsylvania 15260, USA}
\newcommand{\Guanajuato}{Campus Le\'{o}n y Campus Guanajuato, Universidad de Guanajuato, Lascurain de Retana No. 5, Colonia Centro, Guanajuato 36000, Guanajuato M\'{e}xico.}
\newcommand{\Athens}{Department of Physics, University of Athens, GR-15771 Athens, Greece}
\newcommand{\Tufts}{Physics Department, Tufts University, Medford, Massachusetts 02155, USA}
\newcommand{\WM}{Department of Physics, College of William \& Mary, Williamsburg, Virginia 23187, USA}
\newcommand{\FNAL}{Fermi National Accelerator Laboratory, Batavia, Illinois 60510, USA}
\newcommand{\Purdue}{Department of Chemistry and Physics, Purdue University Calumet, Hammond, Indiana 46323, USA}
\newcommand{\MCLA}{Massachusetts College of Liberal Arts, 375 Church Street, North Adams, MA 01247}
\newcommand{\UMD}{Department of Physics, University of Minnesota -- Duluth, Duluth, Minnesota 55812, USA}
\newcommand{\Northwestern}{Northwestern University, Evanston, Illinois 60208}
\newcommand{\UNI}{Universidad Nacional de Ingenier\'{i}a, Apartado 31139, Lima, Per\'{u}}
\newcommand{\Rochester}{University of Rochester, Rochester, New York 14627 USA}
\newcommand{\Austin}{Department of Physics, University of Texas, 1 University Station, Austin, Texas 78712, USA}
\newcommand{\USM}{Departamento de F\'{i}sica, Universidad T\'{e}cnica Federico Santa Mar\'{i}a, Avenida Espa\~{n}a 1680 Casilla 110-V, Valpara\'{i}so, Chile}
\newcommand{\Geneva}{University of Geneva, 1211 Geneva 4, Switzerland}
\newcommand{\Chicago}{Enrico Fermi Institute, University of Chicago, Chicago, IL 60637 USA}
\newcommand{\hired}{}
\newcommand{\OregonState}{Department of Physics, Oregon State University, Corvallis, Oregon 97331, USA}
\newcommand{\oxford}{Oxford University, Department of Physics, Oxford, United Kingdom}
\newcommand{\umiss}{University of Mississippi, Oxford, Mississippi 38677, USA}
\newcommand{\upenn}{Department of Physics and Astronomy, University of Pennsylvania, Philadelphia, PA 19104}
\newcommand{\AMU}{AMU Campus, Aligarh, Uttar Pradesh 202001, India}
\newcommand{\wroclaw}{University of Wroclaw, plac Uniwersytecki 1, 50-137 Wroclaw, Poland}
\newcommand{\Mohali}{IISER, Mohali, Knowledge city, Sector 81, Manauli PO 140306}


\author{T.~Le}                            \affiliation{\Tufts}  \affiliation{\Rutgers}
\author{F.~Akbar}                         \affiliation{\AMU}
\author{L.~Aliaga}                        \affiliation{\WM}  \affiliation{\PUCP}
\author{D.A.~Andrade}                     \affiliation{\Guanajuato}
\author{M.~V.~Ascencio}                   \affiliation{\PUCP}
\author{A.~Bashyal}                       \affiliation{\OregonState}
\author{A.~Bercellie}                     \affiliation{\Rochester}
\author{M.~Betancourt}                    \affiliation{\FNAL}
\author{A.~Bodek}                         \affiliation{\Rochester}
\author{J.~L.~Bonilla}                    \affiliation{\Guanajuato}
\author{A.~Bravar}                        \affiliation{\Geneva}
\author{H.~Budd}                          \affiliation{\Rochester}
\author{G.~Caceres}                       \affiliation{\CBPF}
\author{T.~Cai}                           \affiliation{\Rochester}
\author{M.F.~Carneiro}                    \affiliation{\OregonState}
\author{D.~Coplowe}                       \affiliation{\oxford}
\author{S.A.~Dytman}                      \affiliation{\Pittsburgh}
\author{G.A.~D\'{i}az~}                   \affiliation{\Rochester}  \affiliation{\PUCP}
\author{J.~Felix}                         \affiliation{\Guanajuato}
\author{L.~Fields}                        \affiliation{\FNAL}  \affiliation{\Northwestern}
\author{A.~Filkins}                       \affiliation{\WM}
\author{R.~Fine}                          \affiliation{\Rochester}
\author{N.~Fiza}                          \affiliation{\Mohali}
\author{A.M.~Gago}                        \affiliation{\PUCP}
\author{H.~Gallagher}                     \affiliation{\Tufts}
\author{A.~Ghosh}                         \affiliation{\USM}  \affiliation{\CBPF}
\author{R.~Gran}                          \affiliation{\UMD}
\author{D.A.~Harris}                      \affiliation{\FNAL}
\author{S.~Henry}                         \affiliation{\Rochester}
\author{S.~Jena}                          \affiliation{\Mohali}
\author{J.~Kleykamp}                      \affiliation{\Rochester}
\author{M.~Kordosky}                      \affiliation{\WM}
\author{D.~Last}                          \affiliation{\upenn}
\author{X.-G.~Lu}                         \affiliation{\oxford}
\author{E.~Maher}                         \affiliation{\MCLA}
\author{S.~Manly}                         \affiliation{\Rochester}
\author{W.A.~Mann}                        \affiliation{\Tufts}
\author{C.~Mauger}                        \affiliation{\upenn}
\author{K.S.~McFarland}                   \affiliation{\Rochester}  \affiliation{\FNAL}
\author{A.M.~McGowan}                     \affiliation{\Rochester}
\author{B.~Messerly}                      \affiliation{\Pittsburgh}
\author{J.~Miller}                        \affiliation{\USM}
\author{J.G.~Morf\'{i}n}                  \affiliation{\FNAL}
\author{D.~Naples}                        \affiliation{\Pittsburgh}
\author{J.K.~Nelson}                      \affiliation{\WM}
\author{C.~Nguyen~}                       \affiliation{\Florida}
\author{A.~Norrick}                       \affiliation{\WM}
\author{Nuruzzaman}                       \affiliation{\Rutgers}  \affiliation{\USM}
\author{A.~Olivier}                       \affiliation{\Rochester}
\author{V.~Paolone}                       \affiliation{\Pittsburgh}
\author{G.N.~Perdue}                      \affiliation{\FNAL}  \affiliation{\Rochester}
\author{M.A.~Ram\'{i}rez}                 \affiliation{\Guanajuato}
\author{R.D.~Ransome}                     \affiliation{\Rutgers}
\author{H.~Ray}                           \affiliation{\Florida}
\author{D.~Rimal}                         \affiliation{\Florida}
\author{D.~Ruterbories}                   \affiliation{\Rochester}
\author{H.~Schellman}                     \affiliation{\OregonState}  \affiliation{\Northwestern}
\author{J.T.~Sobczyk}                     \affiliation{\wroclaw}
\author{C.J.~Solano~Salinas}              \affiliation{\UNI}
\author{H.~Su}                            \affiliation{\Pittsburgh}
\author{M.~Sultana}                       \affiliation{\Rochester}
\author{V.S.~Syrotenko}                   \affiliation{\Tufts}
\author{E.~Valencia}                      \affiliation{\WM}  \affiliation{\Guanajuato}
\author{M.Wospakrik}                      \affiliation{\Florida}
\author{B.~Yaeggy}                        \affiliation{\USM}
\author{L.~Zazueta}                       \affiliation{\WM}


\collaboration{The MINERvA Collaboration}\ \noaffiliation
\date{\today}

\begin{abstract}
The antineutrino scattering channel $\anumu \,\text{CH} \rightarrow \mu^{+} \,\pi^{-} \,X$(nucleon(s)) is analyzed
in the incident energy range 1.5 to 10 GeV using the MINERvA detector at Fermilab. 
Differential cross sections are reported as functions of $\mu^{+}$ momentum and production angle, 
$\pi^{-}$ kinetic energy and production angle, and antineutrino energy and squared four-momentum transfer. 
Distribution shapes are generally reproduced by
simulations based on the GENIE, NuWro, and GiBUU event generators, however GENIE (GiBUU) 
overestimates (underestimates) the cross-section normalizations by 8\% (10\%).  
Comparisons of data with the GENIE-based reference simulation probe conventional treatments of cross
sections and pion intranuclear rescattering.    The distribution of non-track vertex energy is used to 
decompose the signal sample into reaction categories, and cross sections are determined for the exclusive reactions
$\mu^{+} \pi^{-} n$ and $ \mu^+ \pi^{-} p$.   A similar treatment applied to the published MINERvA sample 
$\anumu \,\text{CH} \rightarrow \mu^{+} \,\pi^{0} \,X$(nucleon(s)) has determined the $\mu^{+} \pi^{0} n$ cross section,
and the latter is used with $\sigma(\pi^{-} n)$ and $\sigma(\pi^{-} p)$ to carry out an isospin decomposition
of $\anumu$-induced CC($\pi$).   The ratio of magnitudes and relative phase for isospin amplitudes $A_{3}$
and $A_{1}$ thereby obtained are:  $R^{\anu} = 0.99 \pm 0.19$ and $\phi^{\anu} = 93^{\circ} \pm 7^{\circ}$.
Our results are in agreement with bubble chamber measurements made four decades ago.
\end{abstract}

\maketitle


\section{Introduction}

An international effort is underway to determine the ordering of
neutrino mass eigenstates, to delimit the amount of charge conjugation plus parity (CP) violation in the
neutrino sector, and to measure the angles that characterize neutrino
flavor mixing.   To achieve the levels of precision that these goals
require, neutrino flavor oscillations must be investigated using
$\anumu$ as well as $\numu$ beams because antineutrino versus neutrino propagation in
matter elicits differences that are highly informative.   Comparisons of 
antineutrino versus neutrino oscillations are best carried out using 
the same long-baseline and source of $\nu$ fluxes.  This general strategy
underwrites the ongoing experimental programs of T2K~\cite{T2K-expt} and NOvA~\cite{NOvA-expt}, and
it strongly shapes the DUNE program~\cite{DUNE-expt}.   In recent times, combined analyses
of $\numu$ and $\anumu$ oscillations have been reported by T2K and 
NOvA, with each experiment restricting to its own data~\cite{T2K-PRL-2018, NOvA-PRD-2018}.   These observations
allow large values for the Dirac CP-violating phase, and they permit the atmospheric mixing angle $\theta_{23}$ 
to have values in either the lower or upper octant, or to coincide with maximal mixing at $45^{\circ}$.
At the present time, an unambiguous picture for the 
neutrino sector continues to elude.   For continued progress,
the details of antineutrino-nucleus scattering must be established
at a level of accuracy that heretofore has not been available.
Such an understanding must encompass $\anumu$ scattering on nuclear
media used in long baseline experiments, of which hydrocarbon is the simplest representative.

There has been a dearth of measurements 
for charged current (CC) single pion production by antineutrino-nucleus scattering 
in the threshold-to-few GeV region of incident
$\anumu$ energy, $E_{\anu}$~\cite{Katori-Martini-2018}.   This work addresses the situation by presenting
detailed measurements of the semi-exclusive antineutrino interaction channel 
\begin{equation}
\label{signal-channel}
  \anumu  + \text{CH}  \rightarrow  \mu^{+} + \pi^{-} + X(\text{nucleon(s))}.
\end{equation}
Here, the hadronic system $X$ may contain any number of protons and neutrons, but no additional mesons.
For the selected events,  $X$ will consist of an interaction neutron or proton, plus
remnant nucleons from breakup of the target nucleus.  

Signal channel \eqref{signal-channel} receives large contributions from two CC exclusive reactions:
\begin{equation}
\label{exclusive-channel-1}
\anumu + \textrm{n} \rightarrow\mu^{+}+\pi^{-}+\textrm{n}, 
\end{equation}
and
\begin{equation}
\label{exclusive-channel-2}
\anumu + \textrm{p} \rightarrow\mu^{+}+\pi^{-}+\textrm{p} .
\end{equation}

The scattering is dominated by interactions within carbon nuclei, however reaction \eqref{exclusive-channel-2} can take 
place on hydrogen as well.    The signal channel is affected by migrations to and from other channels
as the result of nuclear medium effects.   For example, intranuclear absorption of $\pi^{-}$ mesons initially created by 
channel \eqref{signal-channel} within carbon nuclei depletes the signal-channel rate 
that would otherwise be obtained if the interactions occurred on free nucleons.  On the
other hand, CC multipion production followed by intranuclear pion absorption gives a
rate enhancement to the observable (out of parent nucleus) final states of channel \eqref{signal-channel} that originates from reactions
that are not as-born CC single $\pi^{-}$ occurrences.   Additionally, charge exchange within the struck nucleus can move events out 
of or into ($\pi^- p \leftrightarrow \pi^0 n$) channel \eqref{signal-channel}.

Channel~\eqref{signal-channel} receives a small contribution from CC coherent 
single $\pi^{-}$ production wherein an incident $\anumu$ scatters from the entire target nucleus:
\begin{equation}
\label{coherent-pion-production}
\anumu + \mathcal{A} \rightarrow \mu^{+} + \pi^{-} + \mathcal{A},
\end{equation}
The cross section for reaction \eqref{coherent-pion-production} on carbon has been previously measured
by MINERvA~\cite{ref:CC-Coh-Minerva, Mislivec-2018}.

The CC interactions that comprise channel~\eqref{signal-channel} 
are of keen interest to the NOvA and T2K analyses of $\anumu$ oscillations, 
since antineutrino CC(1$\pi$) channels give significant event rates in the one to few-GeV region 
of $E_{\anu}$.   This $E_{\anu}$ range is affected by $\nue$ flavor appearance and $\numu$ flavor disappearance over
the long baselines used by these experiments, and this will also be the case for the next-generation
long-baseline oscillation experiments, DUNE and Hyper-Kamiokande~\cite{HyperK-expt}.  

The analysis presented here obtains differential cross sections for channel~\eqref{signal-channel} 
that characterize the kinematics of both the final-state
$\mu^{+}$ and the produced $\pi^{-}$.  
These differential cross sections complement and extend MINERvA's 
previously reported measurements of CC pion production on hydrocarbon.
The latter measurements include $\anumu$-induced 
CC(1$\piz$) production~\cite{Trung-pion, Carrie-pion}, 
and $\numu$-induced CC($\pi^{+}$) and CC($1\piz$) 
production~\cite{Brandon-pion, Carrie-pion, Altinok-2017}.

\subsection{$\anumu$-CC($\pi^{-}$) measurements and phenomenology}

Current knowledge concerning channel \eqref{signal-channel} and reactions 
\eqref{exclusive-channel-1} and \eqref{exclusive-channel-2} is based on
bubble chamber antineutrino experiments of the 1970s and 1980s.    
Cross sections for reactions \eqref{exclusive-channel-1} and \eqref{exclusive-channel-2}
taking place in propane + freon mixtures
were obtained in the few-GeV region ($<E_{\anu}>$ = 1.5 GeV) 
using Gargamelle~\cite{Gargamelle-1979, Pohl-nu-cc-pi-1979} and over the range 3 to 30 GeV 
using SKAT\cite{SKAT-1989}.   Investigations of both reactions
for incident $\anumu$ energies exceeding 5 GeV were carried out using large 
deuterium-filled bubble chambers~\cite{Barish-antinu-1980, BEBC-D-antinu-1983, Allasia-1990}, 
and reaction \eqref{exclusive-channel-2} was studied over the range 
5 $< E_{\anu} <$ 120 GeV using BEBC with a hydrogen fill~\cite{Allen-NP-1986}.
The relative contributions from baryon resonances  
was found to be rather different in the two exclusive reactions:
Reaction~\eqref{exclusive-channel-1} is an I\,=\,3/2 channel in which
production of the $\Delta^{-}(1232)$ resonance plays a major role, while 
~\eqref{exclusive-channel-2} contains I = 1/2 as well as I = 3/2 amplitudes.  
For reaction~\eqref{exclusive-channel-2} at multi-GeV incident energies, production of 
I = 1/2 baryon resonances -- the $N^*$(1520), $N^*$(1535),
and higher mass $N^*$ states -- was reported to be comparable to $\Delta$ production.

Event samples recorded by the bubble chamber experiments were often limited 
to a few hundred events.   The present work
benefits from higher statistics afforded by MINERvA exposures to the 
intense, low energy NuMI antineutrino beam at Fermilab~\cite{NuMI-Beam-2016}.
Furthermore it is carried out for an $E_{\anu}$ range 
that intersects the T2K range and spans the ranges of NOvA, and DUNE, and it utilizes a hydrocarbon target medium
whose nuclear composition is very close to 
that of the NOvA detectors while also approximating the target media used by T2K.

Neutrino experimentation has benefitted from a recent surge in theoretical studies that address
neutrino-induced CC(1$\pi$) production~\cite{Katori-Martini-2018}.    
On the other hand, antineutrino CC(1$\pi$) production on nuclei 
has received a relatively limited treatment~\cite{Paschos-Schalla-arXiv-1209,
Mosel-PRC91-2015, MMB-2016}, although the situation 
is improving~\cite{Ghent-Model-I, Ghent-Model-II, Angular-2018, Kabirnezhad-2018}.
To date, $\anumu$-induced pion distributions 
in momentum and in production angle have been predicted for MINERvA based upon the 
GIBUU neutrino generator~\cite{Mosel-PRC91-2015}, and cross sections on nuclei for $0.5 \leq E_{\anu} \leq 3.0$ GeV 
have been predicted for reactions~\eqref{exclusive-channel-1} and \eqref{exclusive-channel-2}~\cite{MMB-2016}.
For the latter two reactions as they occur on quasi-free nucleons, the classic 
Rein-Sehgal treatment~\cite{Rein-Sehgal-1981, Rein-Z-1987} 
provides a phenomenological framework which is assimilated
into several of the current neutrino event generators.

\section{Overview of Data and Analysis}
\subsection{Detector, Exposure, and $\anu$ Flux}
\label{subsec:B-D-E}

Interactions of muon antineutrinos from the NuMI beam at Fermilab~\cite{NuMI-Beam-2016} 
were recorded in the fine-grained plastic-scintillator tracking detector of MINERvA~\cite{minerva-NIM-2014, minerva-NIM-2015}. 
The detector's central tracking region is surrounded by electromagnetic and hadronic calorimeters, providing event containment.  
The magnetized MINOS near detector, located 2\,m downstream of MINERvA, serves as the muon spectrometer~\cite{minos-NIM-2008}.  
The analysis uses a hexagonal cross section fiducial volume of 2.0 m minimal diameter 
that extends 2.4 m along the beam direction and has a mass of 5570 kg.   
The fiducial volume consists of 112 planes composed of polystyrene scintillator strips 
with triangular cross sections of 1.7\,cm height, 3.3\,cm width,
laid transversely to the detector's horizontal axis.    
The planes of the central tracking region (``tracker")
 are configured in modules with two planes per module; an air gap of 2.5\,mm separates each module.  
The detector horizontal axis is inclined at 3.34$^{\circ}$ relative to the beam direction.
 Three scintillator-plane orientations, at 0$^\circ$ and $\pm 60^\circ$  relative to the detector 
vertical axis, provide  X, U, and V ``views'' of interactions in 
the scintillator.  The planes alternate between UX and 
VX pairs,  enabling 3-D reconstruction of interaction vertices, charged tracks, and 
electromagnetic showers.  
Surrounding the downstream and outer side surfaces of the central tracker are the tracking layers of the
electromagnetic and hadronic calorimeters, designated ECAL and HCAL respectively.   The ECAL regions lie within the 
HCAL and are in contact with the outer layers of the central tracker.   The ECAL is of similar construction to
the central tracker but includes a 0.2\,cm (0.35 radiation length) lead sheet in front
of every plane of scintillator.   The HCAL surrounds the ECAL; it consists of alternating layers of scintillator and 2.54\,cm
thick steel plates.    The readout electronics have a timing resolution 
of 3.0\,ns for hits of minimum ionizing particles~\cite{Marshall-2016}, enabling 
efficient separation of multiple interactions within a  
single 10\,$\mu$s beam spill.

A $\mu^{+}$ that exits the downstream surface
of MINERvA is tracked by the magnetized, steel-plus-scintillator planes 
of MINOS, and its momentum and charge are measured.  
Trajectories of individual muons traversing the two detectors are matched together
by correlating the positions, angles, and timings of track segments in each detector.

The data were taken between September 2010 and May 2012 using the 
low-energy NuMI mode, which produces a wide-band beam with antineutrino energies 
extending from 1\,GeV to greater than 20 GeV and a peak energy of 
3\,GeV.   The polarity of current in the magnetic horns in the 
beamline was set to focus $\pi^-$ mesons, providing a $\bar{\nu}_\mu$ enhanced
flux with an exposure of $1.06 \times 10^{20}$ protons on target (POT).

The $\anumu$ flux is calculated using a detailed simulation
of the NuMI beamline based on GEANT4~\cite{Geant4-2003, Allison-2006} v9.2.p03 with the FTFP\_BERT physics list.
The simulation is constrained using proton-carbon yield 
measurements~\cite{Alt-NA49-2007, Barton-PRD-1983, Lebedev-Thesis} together with
more recent thin-target data on hadron 
yields~\cite{NuMI-Flux-Aliaga-2016}.   A further constraint 
is derived using the $\nu + e^{-}$ scattering rate observed by MINERvA~\cite{Park-PRD-2016}. 
Additional details as pertain to the
antineutrino exposures of this work can be found in Ref.~\cite{Patrick-2018}.

\subsection{Neutrino interaction modeling}
\label{subsec:II-B}

The reference Monte Carlo (MC) simulation used by this analysis is built upon
the GENIE 2.8.4 neutrino event generator~\cite{Andreopoulos-NIM-2010}. 
The rendering of antineutrino-nucleus interactions is based upon the same GENIE models 
described in Ref.~\cite{Patrick-2018}.   Additional details concerning GENIE modeling of
CC($\pi$) channels are given in MINERvA publications~\cite{Brandon-pion, Carrie-pion, Altinok-2017}.
Recent developments in neutrino phenomenology motivate certain augmentations to GENIE
that are implemented via event reweighting and by adding a simulated sample 
of quasielastic-like 2-particle 2-hole (2p2h) events~\cite{Rodriques-2p2h-2016}.   
The refinements (described below) are very similar to those used in the reference simulations 
of recent, published MINERvA 
measurements~\cite{Mislivec-2018, Altinok-2017, Patrick-2018, Ren-PRD-2017, Betancourt-2017, Gran-lmt-2018, X-Lu-PRL-2018, Ruterbories-PRD-2019}.
Importantly, all refinements to the GENIE-based MC used here (version designation MnvGENIE v1.2) 
were decided prior to the present work, and the data analyzed here were not used in the GENIE tuning.

In brief, the struck nucleus is treated as a relativistic Fermi gas augmented 
with a high-momentum tail that accounts for short-range correlations~\cite{Bodek-Ritchie-1981}.
Antineutrino-induced pion production arises from interaction with single nucleons and 
proceeds either by baryon-resonance excitation (RES) or by non-resonant Deep Inelastic Scattering (DIS).  
Simulation of baryon resonance pion production is based upon 
the Rein-Sehgal model~\cite{Rein-Sehgal-1981}, updated with 
modern baryon-resonance properties~\cite{PDG-2012}.   
Decays of baryon resonances produced by antineutrinos are generated isotropically
in their rest frames.  Interference among baryon-resonance amplitudes is assumed to be absent.

Concerning non-resonant single pion production, the Rein-Sehgal formalism is not used.   Instead, the
rate of non-resonant pion production is assigned according to the formalism 
of Bodek-Yang~\cite{Bodek-2005-PS} with parameters adjusted 
to reproduce electron and neutrino scattering measurements over the invariant hadronic mass range 
$W < 1.7$~GeV~\cite{Gallagher-2006,Wilkinson-PRD-2014, Rodrigues-EurPhys-2016}.
The total charge of non-resonant pion-nucleon states is constrained by charge conservation.  For antineutrino CC interactions,
if the final-state pion-nucleon total charge is -1, then the particle content is always $\pi^-$n.   
But if the total charge is zero, then the particle content is assigned to be
$\pi^-$p or $\piz$n with probability 2/3 or 1/3 respectively.

An accurate accounting of intranuclear final-state interactions (FSI) for pions and nucleons 
is important for this analysis.   This is because of the large 
pion-nucleon cross sections that occur in the vicinity of  $\Delta$-resonance excitation.    
The GENIE-based simulation however, does not invoke a microscopic cascade involving formation, propagation, 
interaction, and medium modification of $\Delta$ states.  Instead it uses an effective particle cascade in which each
final-state pion or nucleon is allowed to have at most one rescattering interaction before being absorbed or exiting the target nucleus. 
The relative probabilities among scattering processes are assigned according 
to pion-nucleus scattering data~\cite{Dytman-2011-CP}.   
This approach is amenable to simple event reweighting, whereas a full particle cascade 
is much more involved because weights need to be varied for every produced hadron.
The effective cascade approach works well with relatively low-A nuclei such as carbon and oxygen.
Its predictions give good descriptions of FSI distortions observed in
pion distributions by MINERvA studies of CC 
single pion production~\cite{Trung-pion, Brandon-pion, Carrie-pion, Altinok-2017}.

For antineutrino CC pion production, a rate reduction scale factor of 0.50$\pm$0.50 has been applied to 
the default GENIE prediction for the nonresonant pion contribution.
Such a reduction has been shown to improve the agreement between GENIE and 
$\numu$-deuterium bubble chamber data~\cite{Wilkinson-PRD-2014, Rodrigues-EurPhys-2016},
and it also improves the data-versus-MC agreement in the present analysis.

Antineutrino quasielastic-like (QE-like) reactions  are minor sources of background
for signal channel \eqref{signal-channel}.
Nevertheless, QE-like rate enhancement induced by 2p2h processes is addressed by
adding 2p2h events to the reference simulation.
Their generation is based on the Valencia model~\cite{Nieves-PLB-2012, Gran-PRD-2013}, 
but with the interaction rate raised in order to match the data rate observed 
in MINERvA inclusive $\numu$ scattering data~\cite{Rodriques-2p2h-2016}.  This tuning of the 2p2h
component gives a prediction that well-describes MINERvA $\anumu$ CC data for both inclusive
low three-momentum transfer~\cite{Gran-lmt-2018} and exclusive zero-pion samples~\cite{Patrick-2018}.
Additionally, kinematic distortions of QE-like events that arise from 
long-range nucleon-nucleon correlations are included in accord with 
the Random Phase Approximation (RPA) calculations given in Ref.~\cite{Nieves-PRC-2004}.   

Simulation of the coherent CC pion-production reaction \eqref{coherent-pion-production}
is based on the Rein-Sehgal model~\cite{ ref:RS_paper_2},  with parameters tuned 
to give agreement with MINERvA measurements for this channel~\cite{Mislivec-2018}.

\subsection{Predictions using NuWro and GiBUU}
\label{subsec:NuWro}

For all differential cross sections measured in this work, comparisons are made to
the predictions of the GENIE-based reference simulation.
Alternate perspectives are provided using the predictions of
NuWro~\cite{ref:NuWro} and of the 2017 release of GiBUU~\cite{GiBUU-2012, GiBUU-website}. 
These are two completely independent event generators whose physics models differ 
in many ways from those of GENIE.  

 In NuWro, $\Delta(1232)$ production is calculated
using the Adler model~\cite{Adler-Annals-1968, Adler-PRD-1975} 
instead of relying on the Rein-Sehgal phenomenology.
The baryon-resonance region extends to $W < 1.6$\,GeV;
nonresonant pion production is added incoherently as a fraction of DIS, 
where DIS is based upon the Bodek-Yang model~\cite{Bodek-2005-PS}.
Hadronic FSI within parent nuclei are fully treated.  NuWro simulates pion and nucleon FSI using the cascade formalism of 
the Salcedo-Oset model~\cite{Salcedo-NP-1988}. 
It also accounts for nuclear-medium modification 
of $\Delta$ states~\cite{NuWro-medium-2013}.

In GiBUU, baryon-resonance production and non-resonant pion production are broken out into their vector and 
axial vector components.  The vector currents are fully determined by electron-nucleus scattering data 
(MAID 2007~\cite{MAID}).   The axial-vector parts are modeled using Partially Conserved Axial Currents (PCAC) 
and a dipole form factor or a modified dipole form in the case of the $\Delta(1232)$~\cite{Leitner-PRC-2006}, 
with an axial-vector mass of 1.0 GeV.  Strengths of the axial-vector parts are set according to pion production data.  Non-resonant
scattering for hadronic masses below the $\Delta$ is treated according to effective field theory.  The nuclear
model of GiBUU uses a relativistic local Fermi gas to characterize the momenta of nucleons bound within
a potential characterized by a realistic density function.   The hadronic FSI treatment is based on relativistic
transport theory~\cite{GiBUU-website}.   The GiBUU version used by this analysis, hereafter referred to as
GiBUU-2017, does not include the CC coherent reaction~\eqref{coherent-pion-production}, and an
estimate of its contribution based upon MINERvA measurements has been added to its predictions.  
Also, the 2017 version does not contain background contributions to $\anumu$ pion production (as are
included in a 2019 release~\cite{GiBUU-website}).

\subsection{Detector calibrations and event isolation}
\label{subsec:detector-response}

The ionization response of the MINERvA detector to muons and charged hadrons is simulated 
using GEANT4~\cite{Geant4-2003, Allison-2006} v4.9.4p02 with the QGSP\_BERT
physics list.   The ionization energy scale is established by requiring 
the simulation to match reconstructed energies deposited by 
through-going muons that have been momentum-analyzed using the 
magnetized tracking volume of MINOS~\cite{minerva-NIM-2014}.   
For muon $dE/dx$ energy loss, this scale is known to within 2\%.   
For hadronic ionization energy deposits (``hits"), 
the energy assigned in reconstruction makes use of 
calorimetric corrections.   The corrections were initially extracted from simulations~\cite{minerva-NIM-2014}
and subsequently refined and validated using measurements obtained with a scaled-down replicate detector
operated in a low-energy particle test beam~\cite{minerva-NIM-2015}. 
The test beam data, in conjunction with in-situ measurements, enable determinations of tracking efficiencies and 
energy responses to charged pions, protons, and electrons, and establish 
the value of Birks' constant that best describes the scintillator's light yield.   

For each 10\,$\mu$s spill window of the NuMI antineutrino beam, ionization hits
in the scintillator are isolated in time using ``time slices" of tens 
to sub-two-hundred nanoseconds.  As a result, each antineutrino event 
is associated with a unique time slice.
Charged particles initiated by an event traverse the scintillator strips of the central tracker, 
and their trajectories are recorded as individual hits with
specific charge content and time of occurrence.   These ionization hits are 
grouped in time, and neighboring hits in each scintillator plane are gathered into ``clusters".    
Clusters having more than 1\,MeV of energy are matched among 
the three views and tracks are reconstructed from them.   The reconstructions 
achieve a position resolution per plane of 2.7\,mm, and a track angular resolution of
better than 10\,mrad in each view~\cite{minerva-NIM-2014}.

\section{Track reconstruction and energy estimation} 
\label{sec:Reco-and-select-1}

A track of a candidate CC interaction in the central tracker
is designated as the final-state $\mu^+$ if it exits MINERvA's downstream
surface and can be matched with a positively-charged track entering the upstream face of MINOS.
Candidate muons are required to have production angles
$\theta_{\mu}< 25^{\circ}$ relative to the beam direction 
to ensure that they propagate through the MINOS magnetized volume.

Muon reconstruction uses the trajectory segments in both MINERvA and MINOS to achieve
a momentum resolution ($\sigma$ of the residual fractional error) that increases gradually 
from 3.6\% below 2 GeV/c to 7.9\% above 6 GeV/c.
With the reconstruction of muon tracks, there is a small mismodeling of the
efficiency for building single trajectories that traverse both MINERvA and MINOS.
This is addressed by applying a downwards correction of $-4.4\%$ ($-1.1\%$) 
to the simulated efficiency for muons of momenta less than (greater than) 3 GeV/c~\cite{Carrie-pion}.
Upon reconstruction of the $\mu^{+}$ track in an event, the primary vertex location
is estimated using the most upstream hit of the muon and a search is made 
for shorter, hadronic tracks associated with the primary vertex.   
Additional tracks that are found are reconstructed and the vertex position is refit.   
Candidate events are required to have primary vertices that occur 
within the central 112 planes of the scintillator tracking region and are located
at least 22\,cm away from any edge of the planes.   These requirements define the 
vertex fiducial volume whose target mass is 5.57 metric tons and contains 3.41 $\times 10^{30}$ nucleons.  

Events with no reconstructed tracks from the primary vertex 
other than the muon are removed from the analysis.   For the remaining events, it is required that
one and only one charged hadronic track accompanies the $\mu^+$.
The latter tracks may initiate secondary interactions 
that appear as ``kinks" along their trajectories.   In order to associate all ionizations 
from secondary scatters with the originating track, searches are made for additional track segments
starting at the endpoints of tracks already reconstructed.   
The pattern of hit ionizations for the hadronic
track is then examined
for compatibility with charged pion and proton hypotheses.  That is, the 
ionization $dE/dx$ profile is compared to profiles for charged pions 
and for protons calculated using the Bethe-Bloch formula, 
and a particle type is assigned according to likelihood ratios.
An event is retained if the non-muon track is identified in this way as being a charged pion.
Based on its ionization, on the constraint of charge conservation, 
and on the apparent absence of a Michel electron from 
$\pi^+$ decay (see below), such a track 
is highly likely (probability $\simeq$ 0.96) to be a $\pi^{-}$.

The pion kinetic energy, $T_{\pi^-}$, is assigned according to total track range, and the distribution of $T_{\pi^-}$ is
subsequently corrected for residual missing energy using an unfolding procedure (see Sec.~\ref{sec:Pion-Kin}).
For event-by-event estimation of $E_{\pi}$ however, energy from range is augmented
by a sum over ionization hits coincident with the event that lie away from but in proximity to the $\pi^{-}$ track.   Such 
hits are reconstructed according to the detector's calibrated calorimetric response and are designated as $E_{\pi}^{calo}$.
Hits that comprise $E_{\pi}^{calo}$ are required to be $> 10$ cm away from the primary vertex and to lie within a radius
of 65\,cm around the endpoint of the $\pi^{-}$ track.  With this search radius, approximately 83\% of off-track pion-induced
ionizations are captured, while $\sim$50\% of final-state nucleon-induced hits are excluded.
In this way, contamination into $E_{\pi}^{calo}$ from neutron scatters is kept to $\le 10$ MeV on average.

In the reactions of channel \eqref{signal-channel}, the kinetic energy carried by nucleons is 
a sizable fraction of the final-state hadronic energy.  For reaction neutrons and for slow protons as well, most of 
this energy is not represented by ionizations produced in the scintillator tracker. 
In particular, secondary scatters of final-state neutrons occasionally give rise 
to localized ionization clusters -- so-called neutron stars or ``N-stars".
N-stars are usually observed at locations remote
from primary vertices by factors of tens to hundreds of centimeters.
Their energy depositions are much smaller than and are not proportional to the kinetic energy of the
scattering neutrons released in antineutrino 
CC interactions~\cite{MINERvA-neutrons-2019}.   Thus final-state N-stars
in MINERvA contain insufficient information to enable 
neutron kinematic energy to be estimated on an event-by-event basis.   Consequently 
this analysis intentionally avoids the use of nucleon-induced ionizations
-- neither neutron stars, nor hits within 10 cm of the primary vertex from slow protons --
in its estimation of event-by-event $E_{\anu}$.   Instead, the analysis assembles all energies
associated with reconstructed tracks and uses them as input for a kinematic
estimation of $E_{\anu}$ as described in Sec.~\ref{sec:Estimate-E-Q-W-signal-sample}.

\section{Sample selection} 
\label{sec:Reco-and-select-2}

Inclusion of events that have three reconstructed tracks ($\mu^{+}  \pi^{-}$ plus proton) 
was initially considered.    The number of 3-track events that pass the above-listed selections (excluding the 2-track topology 
requirement) is 110 events;  the estimated signal purity of this subsample is 55\%.   Unfortunately, the presence of an additional 
track from the primary vertex gives rise to erroneous event reconstruction and introduces multipion background processes 
that are difficult to constrain.   A full accounting of these aspects would introduce complications into the analysis while 
contributing little of added value.   Consequently the selected sample of this analysis is, very intentionally, 
restricted to two-track topologies, and the low-statistics 3-track subsample is excluded.

Cuts are imposed to ensure accurate interpretation of the event topology and to 
minimize background contamination.   For the reconstructed pion, the start point is required to lie
within 6\,cm of the primary vertex.   This selection ensures proximity to the vertex while allowing a
single hit to be missed, as can happen with a track whose production angle exceeds $60^{\circ}$.
Track reconstruction includes a fit-to-vertex step that ensures a degree of alignment.
On the other hand, selected events must be devoid of ``non-vertex tracks''
whose initial hit is displaced radially by more than 6\,cm from the vertex.    Candidate events may have
ionization hits that do not belong to the primary $\mu^{+}$ and $\pi^{-}$ tracks, provided that they 
are not part of a non-vertex track or of a ``line segment" -- the latter being a reconstructed cluster of hits
that spans four or more contiguous planes.
The $\pi^{-}$ tracks of candidate events are required to stop
in either the scintillator-tracking or ECAL regions of the central tracker.   This requirement is needed to
ensure that particle identification based on $dE/dx$ and kinetic energy reconstruction based on 
range are done reliably.    To this end, $\pi^{-}$ endpoints are required to lie in a volume
of hexagonal cross section surrounding the spectrometer's central axis.  
An apothem of 1\,m is chosen so that all stopping points lie~$\geq$\,15\,cm inside the tracker's outer surfaces.
Variation of this cut by $\pm 5$\,cm results in changes to differential cross sections that lie well within the statistical uncertainties.

The signal channel~\eqref{signal-channel} involves the production of one and only one $\pi^-$ meson.
To eliminate backgrounds that give $\pi^+$ mesons, 
the regions surrounding primary vertices and around track endpoints 
are examined for occurrences of Michel electrons from decays of 
stopped $\pi^{+}$ tracks: $\pi^{+}\rightarrow\mu^{+}\rightarrow e^{+}$.    
Such decays give low-energy ($ \le100$ MeV) EM showers that appear 
later than the candidate-event time by 0.5 to 16\,$\mu$s.    
Events accompanied by a Michel-electron candidate are removed.

\begin{figure}
\begin{center}
\includegraphics[width=8.5cm]{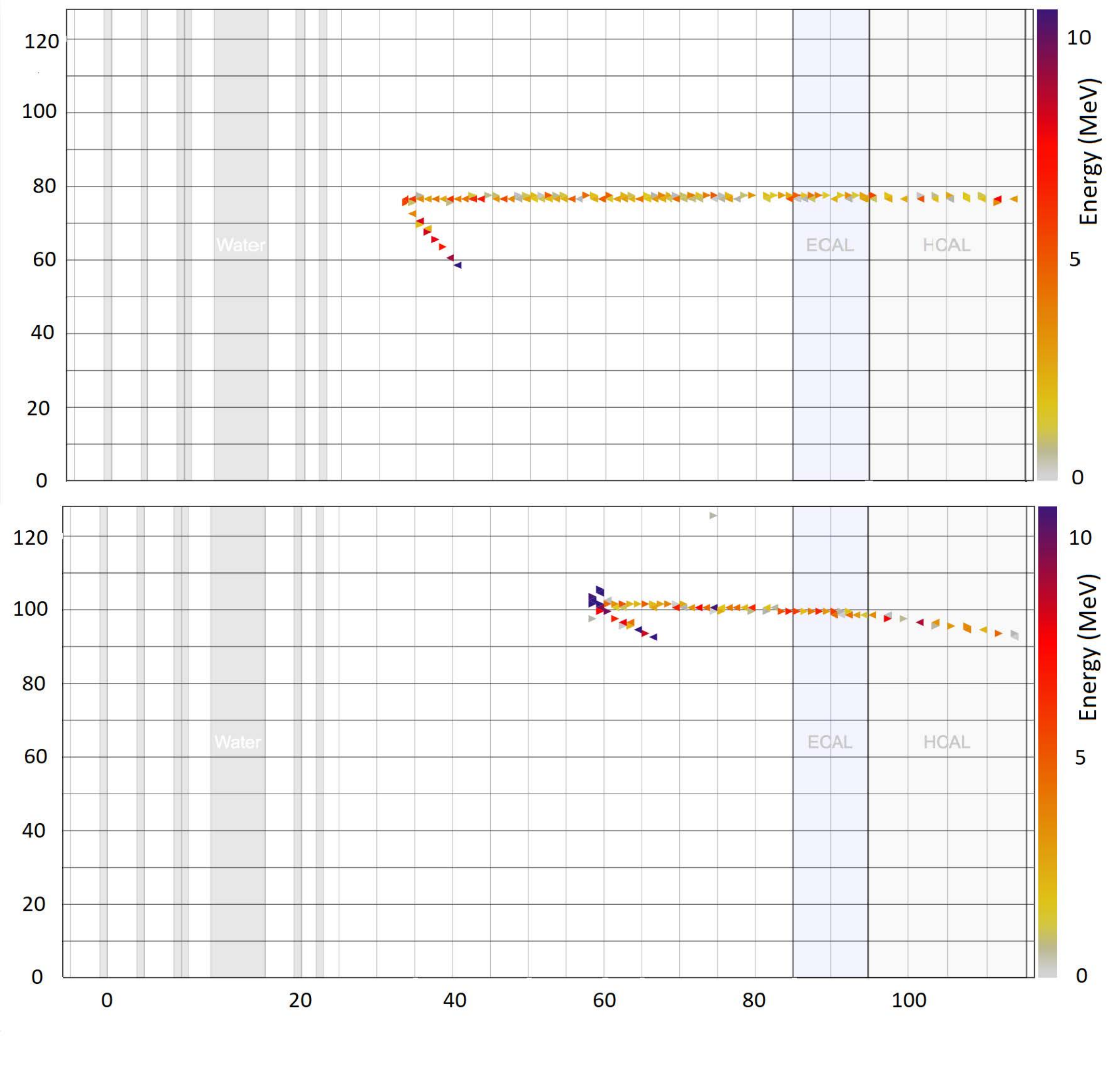}
\caption{Data candidates for signal channel \eqref{signal-channel}.
For each event, the $\anumu$ entered from the left and interacted within 
the central scintillator, yielding a $\mu^{+}$, a charged pion, originating from 
a primary vertex that is devoid of (upper panel) or else has (lower panel) 
additional ionization hits nearby.    Horizontal and vertical axes show module 
and strip numbers respectively.   The right-side linear scale (color online) shows energy deposited in the strips.} 
\label{Fig01}
\end{center}
\end{figure}

Figure~\ref{Fig01} shows two data events from the candidate sample.
Each interaction occurred in the central tracker and is displayed here
in an X-view, looking down at the detector, using the Arachne event viewer~\cite{Arachne}.   
The final-state muons traverse the scintillator planes of the tracker,  
ECAL, and HCAL regions and exit downstream.  These muons give 
matches (spatially and in-time) to $\mu^{+}$ tracks reconstructed 
in the magnetized MINOS detector.    In each event the $\mu^{+}$ is accompanied by 
a charged pion that ranges to stopping.    The pions of
the two events have kinetic energies of 118 MeV (upper panel)
and 173 MeV (lower panel) and are fairly typical of pions in the candidate sample.   

The event of the upper panel is devoid of extra hits around the vertex; the only ionizations are
those on the muon and pion tracks.   Candidate events may have additional hits arising, e.g., 
from inelastic scatters of $\pi^{-}$ tracks or from isolated neutron hits.  More interesting 
are additional hits in the vicinity of the vertex.   Such an occurrence is illustrated by the
event shown in the lower panel of Fig.~\ref{Fig01}.   It has
a pair of extra, heavily-ionized hits -- a pattern that likely originates
from a stopping proton.  These two events indicate how distinctions based
on extra energy at primary vertices can be used to statistically decompose the
signal channel \eqref{signal-channel} into 
exclusive reactions, among which reactions \eqref{exclusive-channel-1} and 
\eqref{exclusive-channel-2} are major contributors.    This line of inquiry 
is pursued in Sec.~\ref{sec:free-nucleon-scattering}.

\section{Kinematic variables and final selections}
\label{sec:Estimate-E-Q-W-signal-sample}
As related above, $p_{\mu}$ is reconstructed
using the muon's curvature and range in MINOS in conjunction with its
$dE/dx$ energy loss as it traverses the MINERvA tracker.
The kinetic energy of the produced $\pi^{-}$, $T_{\pi}$, is assigned using track range.
In traversing MINERvA's hydrocarbon medium however, negative pion tracks can 
undergo inelastic scattering or can be terminated by charge exchange or nuclear 
absorption; consequently track range tends to give an underestimate of  
true pion energy.    To better estimate $E_{\pi}$ of individual events,
the calorimetric energy of ionization hits coincident with an event and
in proximity to the $\pi^{-}$ endpoint (see Sec.~\ref{sec:Reco-and-select-1}) is added to
$T_{\pi}$:  $E_{\pi}$ =  $T_{\pi} + E_{\pi}^{calo}$.    Then the initial direction of the $\pi^{-}$ track, 
together with $|\vec{p}_{\pi}| = \sqrt{E_{\pi}^{2} - m_{\pi}^2}$, establishes the pion 3-vector.

The incident antineutrino energy $E_{\anu}$ is estimated on the basis of the kinematics 
of exclusive CC($\pi$) reactions where the struck nucleon is assumed to be at rest.
Under this approximation, 
the incident antineutrino energy $E_{\anu}$ is calculated according to the relation
\label{kinematic-formula}
\[ \begin{split}
\label{eq:neutrino-energy-estimate}
\nonumber
&E_{\anu}^{\text{CC}(\pi)} \,= \\
&\frac{m_{\mu}^2 + m_{\pi}^2 - 2 m_{N}E_b + E_b^2 -2 m_{N_{b}} (E_{\mu} + E_{\pi}) + 2 \mathcal{P}_{\mu}\cdot \mathcal{P}_{\pi}}
 {2\,[ E_{\mu} + E_{\pi} - |\vec{p}_{\mu}|\cos\theta_{\nu,\mu} - |\vec{p}_{\pi}|\cos\theta_{\nu,\pi} - m_{N_{b}} ]} .
\end{split} \]
Here, the 4-vector product in the numerator is 
$\mathcal{P}_{\mu} \cdot \mathcal{P}_{\pi} = E_{\mu}E_{\pi} - \vec{p}_{\mu} \cdot \vec{p}_{\pi}$,
and $m_{N_{b}}$ denotes the nucleon mass reduced by the binding energy, $E_{b}$, of the initial state nucleon:
$m_{N_{b}} = ( m_{N} - E_{b} )$.    A value of 30 MeV is assigned to $E_{b}$ based 
on electron scattering data~\cite{Moniz-1971, Orden-1981}.  

The kinematic constraint for CC($\pi^{-}$) channels utilized here is a modestly-refined version of the formula used previously
by MiniBooNE in analysis of $\numu$-CC($\pi^{+}$) scattering~\cite{MiniBooNE-piplus-2011}.   In essence, the formula
accounts for invisible nucleon kinetic energy by requiring the vector momenta of final-state particles to balance with
respect to directions transverse to the $\anumu$ beam.

With event $E_{\anu} = E_{\anu}^{\text{CC}(\pi)}$ determined as above,
the nucleon $T_{N}$ of each event (that is, the estimated kinetic energy of the interaction nucleon, neglecting
Fermi motion and nuclear breakup contributions) can be inferred:
$T_{N} =  E_{\anu} - (E_{\mu} + E_{\pi} + E_{b}).$
The shape of the data $T_{N}$ spectrum obtained in this way peaks at 60 MeV
and falls away approximately exponentially, reaching negligible rate by 1.0 GeV.
Since the reference MC reproduces the derived spectral shape to within 17\% over the 
full data range, it is reasonable to query the underlying simulation for some
rough characterizations of neutron production:   According to the MC, the average $T_{N}$ per event
is $\sim$113 MeV for the selected sample.   The average exhibits a
linear correlation with incident $\anumu$ energy, varying from 
75 MeV for $E_{\anu}$ below 3 GeV, to 150 MeV
for $E_{\anu} = 9$ GeV.   Final-state $T_{N}$ is estimated to account for
 2.9\% of event $E_{\anu}$ on average.

For $E_{\anu}$ and for all other measured quantities in this work, the resolution is calculated as
the r.m.s. width of the fractional residual error.   The resolution for $E_{\anu}$ is 9.5\%.   With
event-by-event estimations of $E_{\anu}$ in hand,
the four-momentum-transfer squared, 
$Q^{2}$, and the hadronic invariant mass, $W$, are then calculated as follows:
\begin{equation}
\label{def-Q2}
Q^2 = -(k-k')^2 =  2E_{\anu} (E_\mu-|\vec{p}_\mu|\cos\theta_{\mu})-m_\mu^2 ,
\end{equation}
and
\begin{equation}
\label{def-W}
 W^2 = (p+q)^2 = m_{N}^2 + 2m_{N}(E_{\anu} - E_\mu)- Q^2 .
\end{equation}
Here, $k$, $k'$, and $p$ are the four-momenta of the incident neutrino, the
outgoing muon, and the struck nucleon respectively, while
$q=k-k'$ is the four-momentum transfer and $m_N$ is the nucleon mass. 

The resolution for the variable $Q^2$ is 0.09 GeV$^2$.
Concerning the hadronic mass $W$, the formula of Eq.~\eqref{def-W} is based on the assumption
that the struck nucleon is initially at rest.   It is therefore useful to distinguish between the 
estimator $W_{exp}$ used by this analysis versus the ``true W" of the reference simulation.  
The analysis estimates the hadronic mass, $W_{exp}$, of each signal event 
using Eq.~\eqref{def-W}.  
The resolution in $W_{exp}$ for this analysis  
is 0.12 GeV (0.17 GeV) for $W_{exp} < 1.4$\,GeV ($W_{exp} > 1.4$\,GeV).

As final selections for the signal sample, reconstructed neutrino energies of selected events are
restricted to the range $1.5\textrm{ GeV}<E_{\anu}<10\textrm{ GeV}$ and an upper bound of 1.8 GeV is placed on $W_{exp}$.
The lower bound on $E_{\anu}$, together with the upper bound on $\theta_{\mu}$ 
(see Sec.~\ref{sec:Reco-and-select-1}), ensures good acceptance for muons to be matched in MINOS, and the
upper bound on $W_{exp}$ mitigates background from CC multipion production.   In summary, three kinematic selections
comprise the signal definition of this analysis:  {\it(i)} $\theta_{\mu} < 25^{\circ}$ for the $\mu^{+}$ track at production,
{\it(ii)} $1.5 < E_{\anu} < 10.0$\,GeV for the antineutrino energy, and 
{\it(iii)} $W_{exp} < 1.8$\,GeV for the hadronic invariant mass.

The analysis signal sample after all selections contains 1606 data events.   
The average selection efficiency is the ratio of selected signal events to total signal events.  
This efficiency, as estimated by the simulation, is 5.8\%.
The sample purity, defined as the number of signal events divided by the number of selected events, is also estimated
using the MC.   The purity is 72\%, implying that 
approximately 1156 of selected data events are actual occurrences of channel~\eqref{signal-channel}.    
The average energy of the $\anumu$ flux over the analyzed $E_{\anu}$ range is 3.5 GeV, while
the average $E_{\anu}$ for the selected signal sample is 3.76 GeV.   That the latter average exceeds the former reflects
the rise in the signal channel cross section with increasing $E_{\anu}$ (see Sec.~\ref{sec:Ev-Q2}).

\begin{figure}
  \begin{center}
  \includegraphics[width=8.8cm]{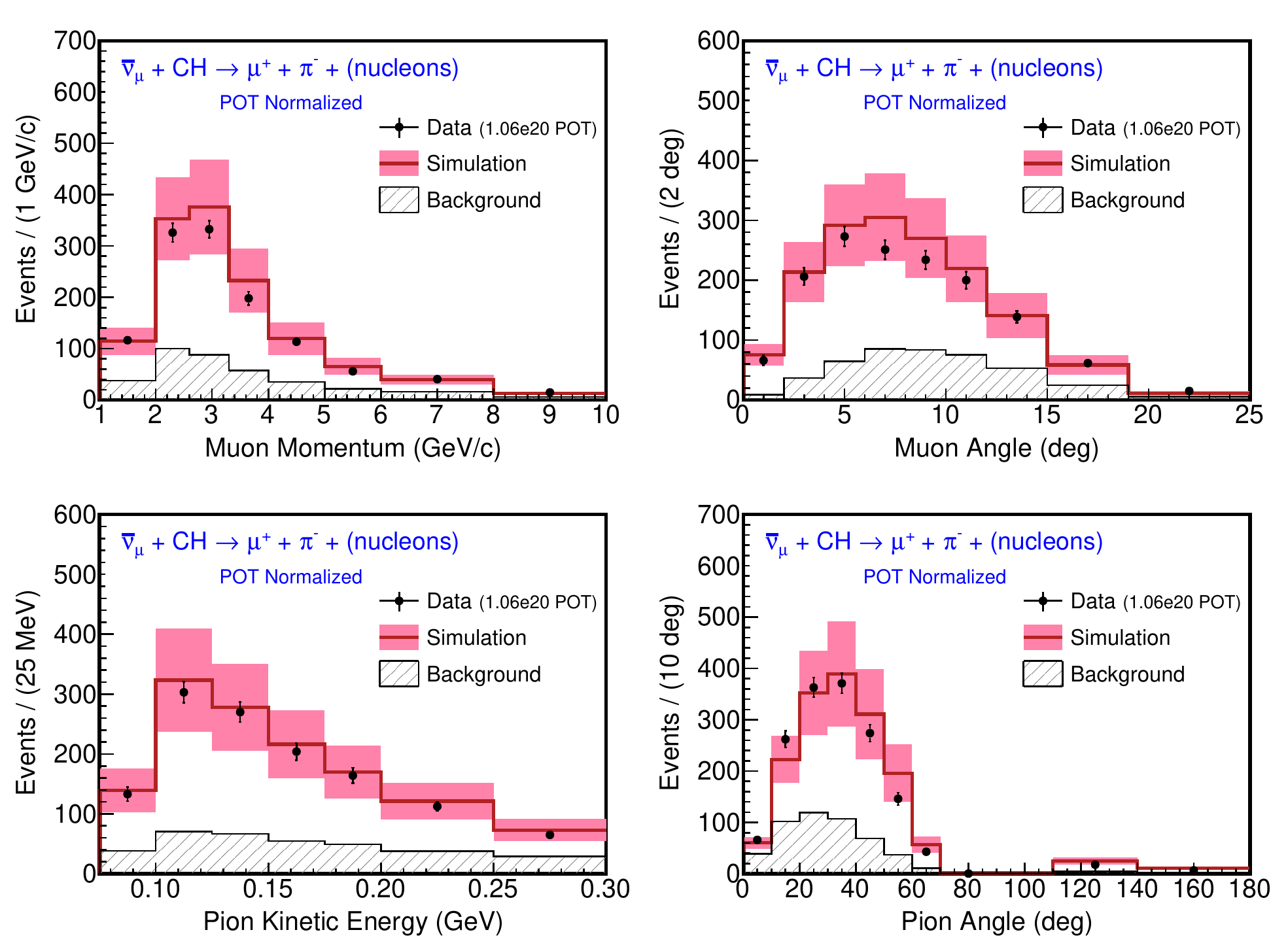}
\caption{Initial data distributions of the selected sample for $\mu^{+}$ and 
$\pi^{-}$ kinematic variables $p_{\mu}$, $\theta_{\mu}$ (upper plots) and 
$T_{\pi}$, $\theta_{\pi}$ (lower plots) compared to the reference MC predictions 
(histograms with systematics error band).  The comparisons 
here are shown before constraining the background (lowermost gray-shade histogram) 
via sideband fitting, and prior to correcting the data for
detector effects.}
\label{Fig02}
\end{center}
\end{figure}

Figure~\ref{Fig02} presents initial comparisons of the selected signal sample to reference MC predictions
using distributions, prior to background subtraction, of directly-measured kinematic variables for final-state 
$\mu^{+}$ and $\pi^{-}$ mesons (upper, lower plots respectively).   
The error bands associated with the MC histograms include uncertainties 
associated with GENIE modeling of both signal and background processes including non-resonant pion 
production as described in Sec.~\ref{subsec:II-B}.
The simulation histograms give
respectable descriptions of the shapes of the data distributions.   
For absolute event rates, however, there is a data-MC offset, with the MC prediction lying above the data in most bins.   This excess rate
predicted by the MC represents an 10\% increase in total event rate compared to the data.  (This initial excess is reduced to 8\% 
by the background constraint of Sec.~\ref{sec:Side-Band-Fit}.)   Nevertheless, the data points are mostly
contained by the $\pm\,1 \sigma$ systematic error band of the MC prediction.
The selected signal sample includes background events, mostly comprised of CC scattering into single-pion or two-pion final states 
that differ from channel \eqref{signal-channel}.   Their contribution is estimated by the reference MC and is shown by
the gray-shade component histograms of Fig.~\ref{Fig02}.    
The overall good agreement between the data and the reference simulation at this stage 
is sufficient to justify its utilization by the analysis to estimate detection efficiencies and to make corrections 
for detector response.
 
\section{background constraint from sideband fitting}
\label{sec:Side-Band-Fit}

The signal sample includes background processes whose final-state particle content upon exit from the target nucleus
is inconsistent with channel \eqref{signal-channel}.    While the reference MC provides
estimates for the rate and kinematic behavior of background events, these estimates come with large uncertainties.
Fortunately, the estimation of background can be greatly constrained by tuning the reference MC to well-describe a background-rich
``sideband sample" whose events have topological and kinematic resemblances to the selected signal events.
A search for a useful sideband was carried out by inspecting samples obtained by
turning off just one selection cut from the ensemble that defines the signal sample.    
Within the full set of cuts there are four specific ones that, when individually reversed, allow
a useful sideband subsample to be defined.   Then, 
by collecting events that pass all signal selections but one, wherein the sole rejection
arises with one of the four specific cuts, a single sideband sample with discriminatory 
power and good statistics is obtained.

The four selection cuts are:  {\it(i)} no reconstructed remote tracks are allowed in the
event, {\it(ii)} all reconstructed line segments must belong to the $\mu^{+}$ or $\pi^{-}$ tracks,
{\it(iii)} the leading hit of the pion track must lie within 6 cm of the 
vertex, and {\it(iv)} the event cannot have a Michel electron.  Each data event of the sideband satisfies all signal selections 
but one, with the excepted selection being one of the four above-listed cuts. 
The sideband sample, assembled in this way, contains  4887 events.

The reference MC is amenable to a simple tuning fit to the sideband;  this situation   
was discerned by comparing the MC predictions to data distributions of the sideband sample using the 
kinematic variables measured by the analysis.   These include
the directly measured variables of $\mu^{+}$ momentum and production angle ($p_{\mu}$ and $\theta_{\mu}$), 
pion kinetic energy and production angle ($T_{\pi}$ and $\theta_{\pi}$),  and the derivative variables
$E_{\anu}$, $Q^{2}$, and $W_{exp}$.    The reference MC was found to describe the shapes of all seven
distributions fairly well, while the absolute rate prediction was higher by $\sim$2\%.

\begin{figure}
  \begin{center}
  \includegraphics[width=8.8cm]{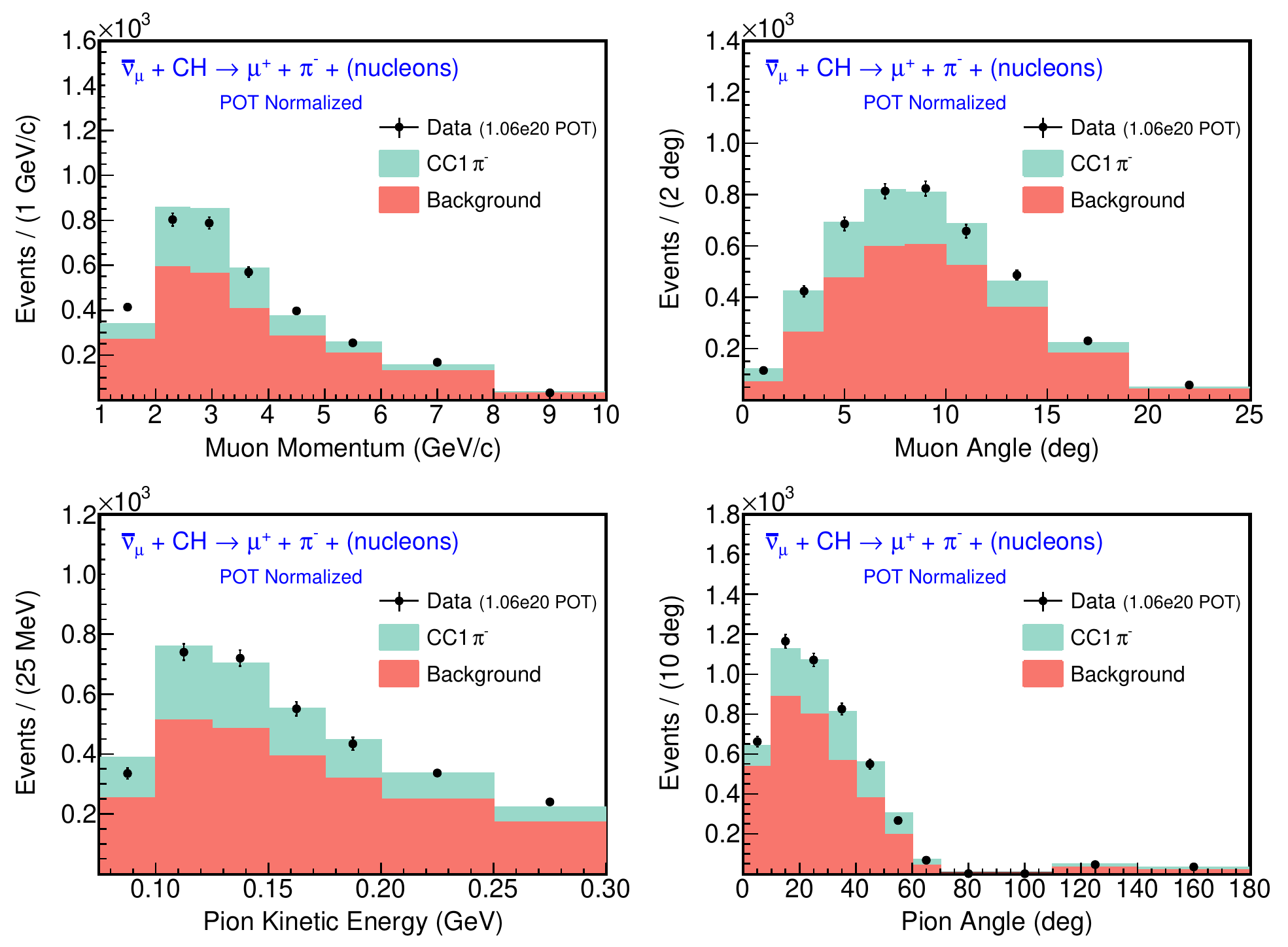}
   \caption{
   Muon and charged pion kinematic distributions for sideband data events (solid points with statistical error bars) compared to the
   reference simulation (histograms) prior to tuning.    The MC describes the shape but slightly overestimates the
   rate of sideband data.   Lower-component histograms (red) show the estimated background content of the sideband. 
   Upper-component histograms (green) depict the signal contamination in the sideband.}
\label{Fig03}
\end{center}
\end{figure}

The initial comparison of the MC with sideband data is displayed in 
Fig.~\ref{Fig03} which shows the sideband distributions for
the kinematic variables of the $\mu^{+}$ and $\pi^{-}$ tracks.
The prediction of the reference MC prior to tuning (histograms) exceeds the sideband data  
in the majority of bins.  Approximately  75\% of the sideband consists of background (lower histograms), originating mostly
from CC RES or non-resonant DIS interaction categories that give rise to multi-pion final states.
 Importantly, the remaining $\sim25\%$ of background is estimated to be
``signal contamination" as shown by the upper component histograms in Fig.~\ref{Fig03}.   This component
of the sideband arises with events that fail the selection criteria as the result of shortfalls in event reconstruction.
Clearly, the presence of signal events in the sideband must be accounted for when fitting the reference MC to match
the sideband distributions.   That said, it is possible to tune the reference MC to match the sideband data distributions
for all seven of the above-listed variables using the iterative procedure described below.

For sideband distributions in each of $p_{\mu}$, $\theta_{\mu}$,
$T_{\pi}$, $\theta_{\pi}$, $E_{\anu}, Q^{2},$ and $W_{exp}$,  the distribution shapes for true background and for signal contamination 
are taken from the MC prediction while the absolute rate normalizations for these two components are treated as parameters in a 
$\chi^{2}$ fit.   Fitting of the MC prediction to the sideband distributions proceeds in two steps, and these are subsequently iterated.  
In the first step, the background normalization for the MC (a single parameter) is allowed to vary in a fit to the seven kinematic distributions
of the sideband data, while the signal contamination normalization is held fixed.
 In the second step, a similar simultaneous fit to the kinematic distributions
of the signal sample is carried out, but with the MC background estimate fixed according to the outcome of step
one, while the normalization of the predicted signal content serves as the fit parameter.   The revised normalizations for MC-estimated
signal and background then serve as input for another two-step fitting sequence.   This two-step fitting of sideband and then signal 
samples is repeated until the background and signal normalizations settle onto stable values.   This fitting procedure converges
with four iterations.  

At this stage the simulation versus data was examined in each bin of the sideband distributions 
for all seven kinematic variable (62 bins) and the verity of predicted rate and shape was evaluated.   
Good agreement was observed overall.   The sole exception was with three contiguous bins 
spanning the peak of the sideband  $W_{exp}$ distribution wherein the MC prediction 
was 1.2-2.5\,$\sigma$ higher than the data.   This mild discrepancy is attributed 
to background events in the simulation, and weights (averaging 0.88) are assigned to MC events 
in the three $W$ bins to bring the simulation closer to the data.   
Incorporation of these weights gives small adjustments ($\leq\,2\%$) to background estimates 
in bins of the other kinematic variables.  An uncertainty of 100\% is assigned 
to the weights and is propagated to the final error budget.   

\begin{figure}
  \begin{center}
  \includegraphics[width=8.8cm]{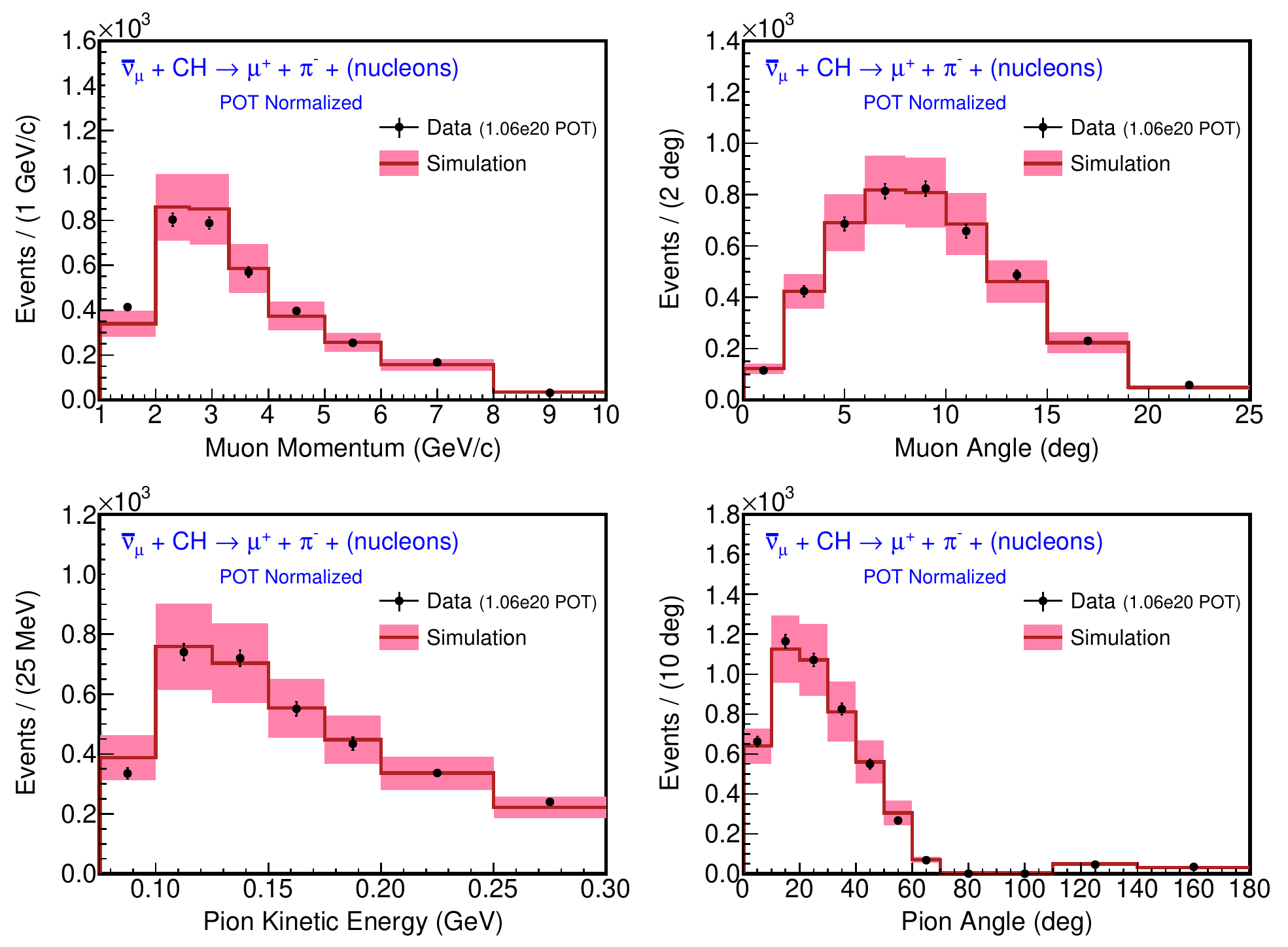}
   \caption{Sideband sample distributions, MC versus data, for muon and pion kinematic variables 
   (upper, lower plots respectively) prior to tuning of background and signal-contamination normalizations.  
   The initial MC predictions and total systematic uncertainties are shown by the histograms
   and shaded error bands.}
\label{Fig04}
\end{center}
\end{figure}

\begin{figure}
  \begin{center}
  \includegraphics[width=8.8cm]{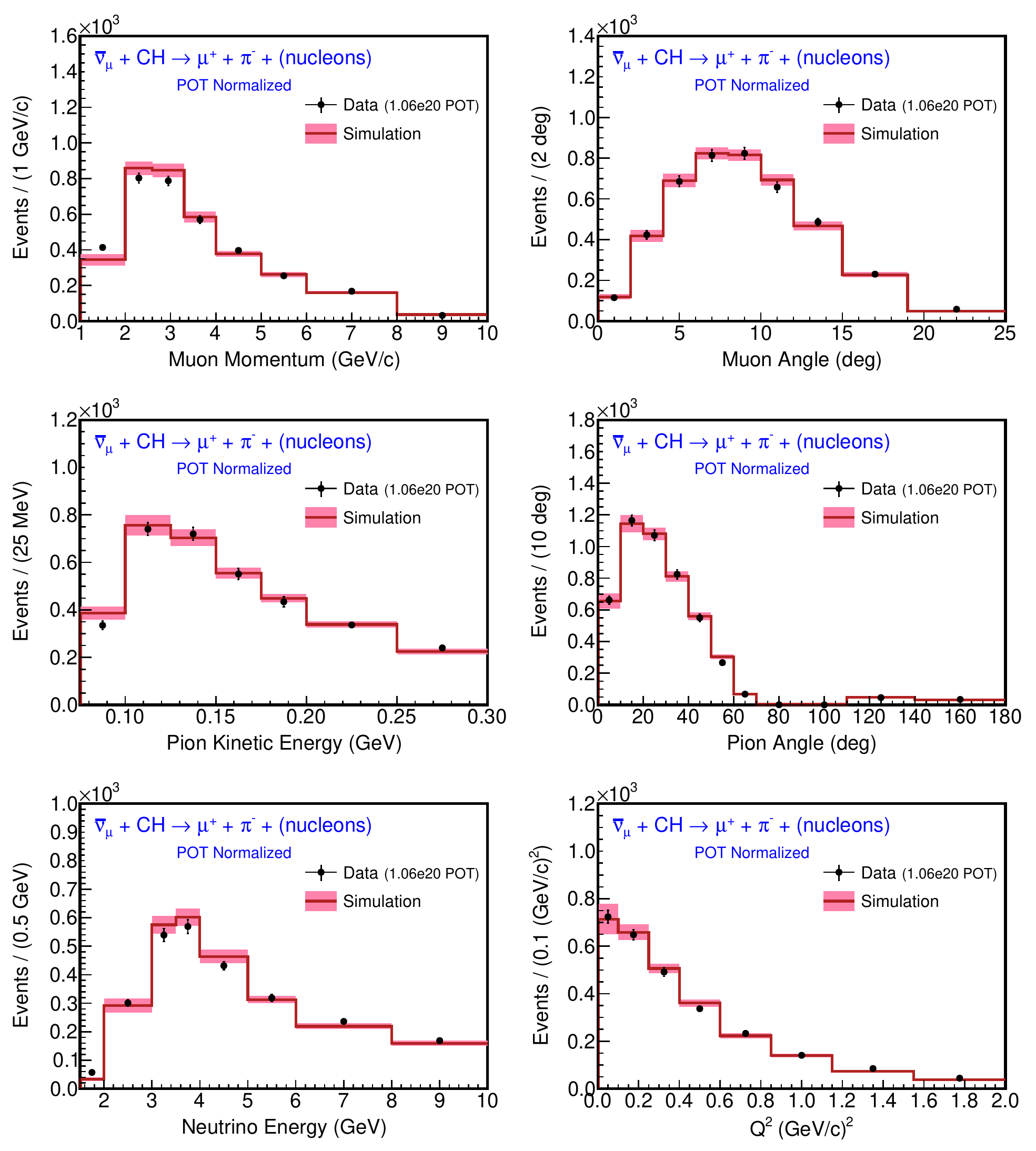}
  \caption{Sideband distributions, MC versus data, for muon and 
  pion variables as in Fig.~\ref{Fig04},  plus distributions for $E_{\anu}$ and $Q^2$.
The MC predictions (histograms with error bands) are shown after the iterative fit of background 
and signal normalizations to seven kinematic distributions of the sideband and signal samples, and  
weight-adjusting the MC in 3 bins of $W_{exp}$.
(see main text).}
\label{Fig05}
\end{center}
\end{figure}

The result of iteratively fitting the background plus signal normalizations and tuning 
the predicted background $W_{exp}$ shape is summarized in 
Figs.~\ref{Fig04}, \ref{Fig05}, and \ref{Fig07}.   Figure~\ref{Fig04} shows 
the sideband distributions of the directly measured muon and pion kinematic variables 
prior to any adjustment.   The reference MC reproduces the distribution shapes quite well, 
with small discrepancies in absolute rate discernible in a few bins.   
The MC predictions, however, have significant flux and GENIE modeling uncertainties 
associated with them, as indicated by the shaded error bands.   
The sideband distributions for these same directly-measured variables after fitting 
and tuning, together with the derivative variables $E_{\anu}$ and $Q^2$, 
are shown in Fig.~\ref{Fig05}.  Here, the match between data points and MC histograms 
is changed slightly by the fitting and tuning procedure.   The main effect is that the fit 
constrains uncertainties associated with event-rate prediction and thus reduces 
the error bands of the tuned MC prediction.  

\begin{figure}
  \begin{center}
  \includegraphics[width=8.9cm]{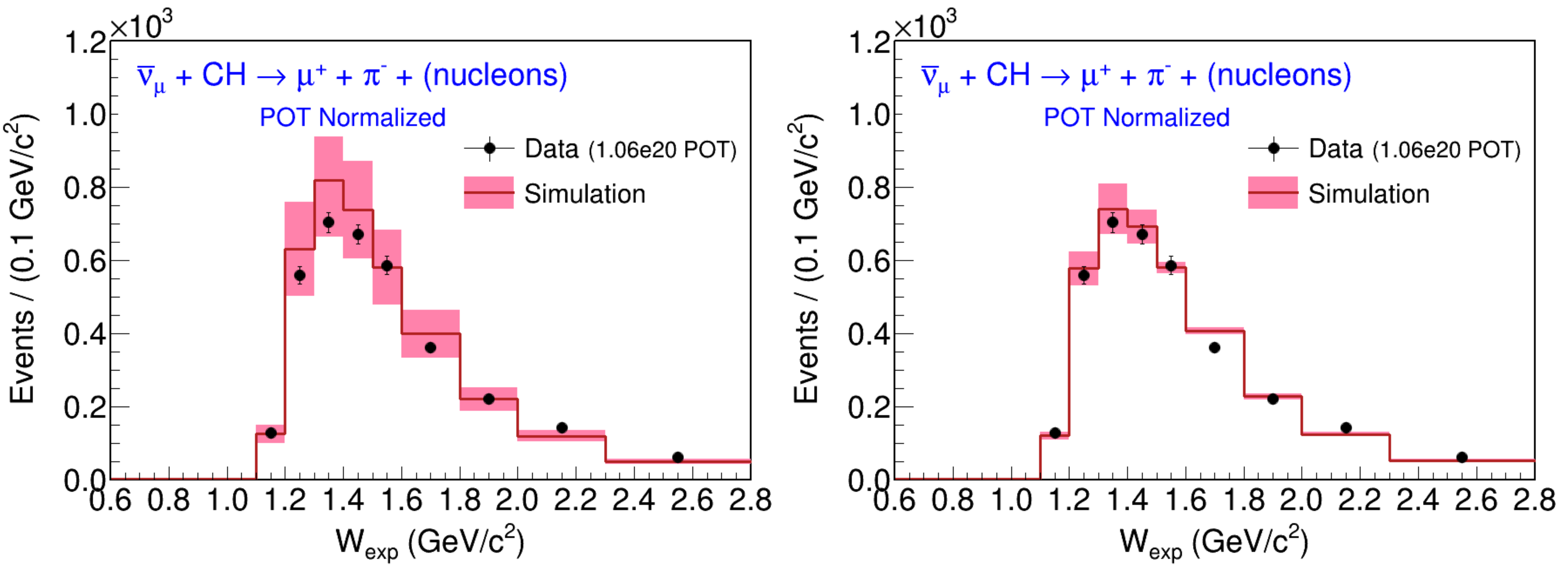}
\caption{Sideband distributions, MC versus data, for estimated hadronic mass $W_{exp}$.
Left-side plot shows sideband $W_{exp}$ prior to any adjustment of the MC.   Right-side plot 
the improved agreement of MC (histogram with error bands) with the data after
fitting of normalizations and weighting of the MC rate through the peak 
($1.2 < W_{exp} < 1.5$ GeV).}
\label{Fig06}
\end{center}
\end{figure}

Figure~\ref{Fig06} shows 
the sideband distribution of the variable least directly measured, 
namely $W_{exp}$, before and after fitting and tuning.    
The initial MC overprediction through the peak region $1.2 < W_{exp} < 1.5$ GeV, 
discernible in Fig.~\ref{Fig06}\,(left), is weight-adjusted 
to give the improved agreement shown in Fig.~\ref{Fig06}\,(right).
The net change to the background normalization from the iterative fit plus shape tuning is an increase of +1\%.
The fit also imposes a 11\% reduction in the estimated signal contamination in the sidebands.    


After tuning the background estimate using the sideband distributions as above,
the reference MC is used to predict the background contribution, $N_{j}^{bkg}$, for
the $j$th bin of any specific distribution of signal-sample events.
The true signal content is then calculated as $(N^{data}_{j} - N^{bkg}_{j})$, where
$N_{j}^{data}$ is the number of data candidates.

\section{Determination of cross sections}
\label{X-sec-calc}
Calculation of the flux-integrated differential cross section per nucleon
for kinematic variable $X$ (such as $p_{\mu}$, $\theta_\mu$, and $Q^{2}$), 
in bins of $i$, proceeds as follows~\cite{Brandon-pion,Trung-pion, Carrie-pion, Altinok-2017}:
\begin{equation}
\label{eq:dif-xsec}
( \frac{d\sigma}{dX} )_{i} =  \frac{1}{\mathcal{T}_{N}\Phi } \frac{1}{\Delta X_i} \frac{1}{\epsilon_{i}}
	\sum\limits_{j} M_{ij} ( N^{data}_{j} - N^{bkg}_{j} ),
\end{equation}
where $\mathcal{T}_{N}$ is the number of target nucleons in the fiducial volume,
$\Phi$ is the integrated flux, $\Delta X_i$ is the bin width,
$\epsilon_{i}$ is the selection efficiency and acceptance.   
The matrix $M_{ij}$ is the unfolding matrix~\cite{{D'Agostini-NIM-1995}}.
It calculates the contribution to true bin $i$ from reconstructed bin $j$, where the $j$th bin contains
$N_{j}^{data}$ number of data candidates and $N_{j}^{bkg}$ number of background events.
Calculation of $\sigma(E_{\anu})_{i}$, the cross section per antineutrino energy
bin $i$, is carried out using
an expression that can be obtained from 
Eq.~\eqref{eq:dif-xsec} by dropping $\Delta X_i$ and changing $\Phi$ to
$\Phi_{i}$, the $\anumu$ flux for the $i$th bin of $E_{\anu}$.

The background-subtracted data is subjected 
to iterative unfolding~\cite{D'Agostini-NIM-1995}.  
The unfolding procedure takes detector resolution smearing into account 
and corrects reconstructed values (j) to true values (i) according to 
mappings, $M_{ij}$, determined by the reference simulation.  
For most of the kinematic variables measured in this work, 
the unfolding matrices are close to diagonal and
the effects of unfolding are minor.   Differences between unfolded distributions
diminish rapidly with consecutive iterations and convergence was achieved
within 3 iterations for $p_{\mu}$, $\theta_{\mu}$, $\theta_{\pi}$, and within 5 iterations
for $E_{\anu}$ and $Q^{2}$.   

Final estimation of $\pi^{-}$ kinetic energy is an exceptional case;  here the unfolding
procedure introduces a significant, necessary correction.    With $T_{\pi}$, visible track range is used
to assign an initial value and it tends to give an underestimate.   This
is because the $T_{\pi}$ of a negative pion, initially produced with several-tens to few-hundreds MeV,
is swept through the $\Delta(1232)$ excitation region as the pion ranges out.
Consequently scattering occurs at elevated rates in modes
that terminate tracks (via charge exchange or absorption) and/or drain away energy via inelastic
transfer to unbinding, recoiling nucleons.   Track ranges thereby tend to be abbreviated, with $T_{\pi}$ being somewhat
underestimated.   Consequently the unfolding procedure requires a relatively large number of iterations in order to converge
to a final result.   The differential cross section
$d\sigma/dT_{\pi^{-}}$ reported in this work (see Sec.~\ref{sec:Pion-Kin}) is obtained using ten unfolding iterations.

For all of the above-mentioned kinematic variables including $T_{\pi}$, the stability of 
unfolded solutions was checked by unfolding ensembles of MC samples representing perturbed variations 
of the initial data distributions.

The bin-by-bin efficiency $\epsilon_{i}$ is estimated using the simulation.
The selection efficiency versus muon momentum, for example, rises 
from 4\% below 2 GeV/c and climbs to 9\% at 4.0 GeV/c, 
as the result of improved tracking acceptance ($\theta_{\mu} < 25^{\circ}$) 
for higher-momentum $\mu^{+}$ tracks 
in the MINOS near detector.   Above 6 GeV, the efficiency gradually diminishes as the
result of the $E_{\nu}$ cut at 10 GeV.
As previously stated, the overall selection
efficiency for signal events is 5.8\%.

The analysis uses current determinations of the integrated 
and differential $\anumu$ fluxes over the $E_{\anu}$ range 1.5 to 10 GeV 
for the NuMI low-energy antineutrino beam mode~\cite{NuMI-Flux-Aliaga-2016}.  
The $\anumu$ flux in bins of $E_{\anu}$ is given in the Supplement~\cite{Supplement}.   
The value for the integrated flux $\Phi$ is 2.00$\times10^{-8}$ $\anumu$/cm$^2$/POT.

\section{Systematic Uncertainties}

Cross-section measurements require knowledge of selection efficiencies, detector acceptance and resolutions,
distribution shapes and normalizations of backgrounds, and the antineutrino flux.    The estimation of each of these quantities
introduces uncertainties.   Many of the sources of uncertainty that affect the present work were encountered 
by previous MINERvA studies of CC($\pi)$ interactions and their treatment 
has been described in publications~\cite{Brandon-pion, Trung-pion, Carrie-pion, Altinok-2017}.   
The systematic uncertainty from the antineutrino flux is described in detail 
in Refs.~\cite{NuMI-Flux-Aliaga-2016, Fields-PRL-2013}.

 The sources of uncertainty can be grouped into six general categories.    
 In Figs.~\ref{Fig07} and \ref{Fig08} of this Section, and in Tables of 
 the Supplement~\cite{Supplement}, the fractional uncertainties 
 for each bin of each measurement are decomposed using 
 these categories.    The first category, designated by ``Detector",  
is assigned to detector response uncertainties arising 
from particle energy scales, particle tracking and detector composition.   
Categories two, three, and four include,
respectively, uncertainties from simulation modeling of neutrino interactions, 
GENIE model uncertainties for FSI involving produced hadrons, and antineutrino flux uncertainties.
These categories are designated as ``X-Sec Model", ``FSI Model", and ``Flux".
Then there are uncertainties that arise with estimation of rate and distribution shapes for the background;
these are compiled in the category labeled ``Bkg Est".
Finally, there are statistical uncertainties that reflect finite sample sizes and the consequent uncertainties that
these generate in the unfolding.   These are included together in the ``Statistical" category.

Systematic uncertainties are evaluated 
by shifting the relevant parameters in the simulation about nominal values
within their $\pm 1\sigma$ bands and producing a new simulated event sample.   
Cross sections are then recalculated using an ensemble of such 
alternate-reality samples, and a covariance matrix is formed from the results.   
The procedure is repeated for each systematic source; 
details are given in Ref.~\cite{Brandon-pion}.   On cross-section plots to follow, 
the error bars shown represent the square roots of covariance diagonal entries.  
The full correlation matrices are given in the Supplement~\cite{Supplement}.

\begin{figure}
  \begin{center}
\includegraphics[width=8.5cm]{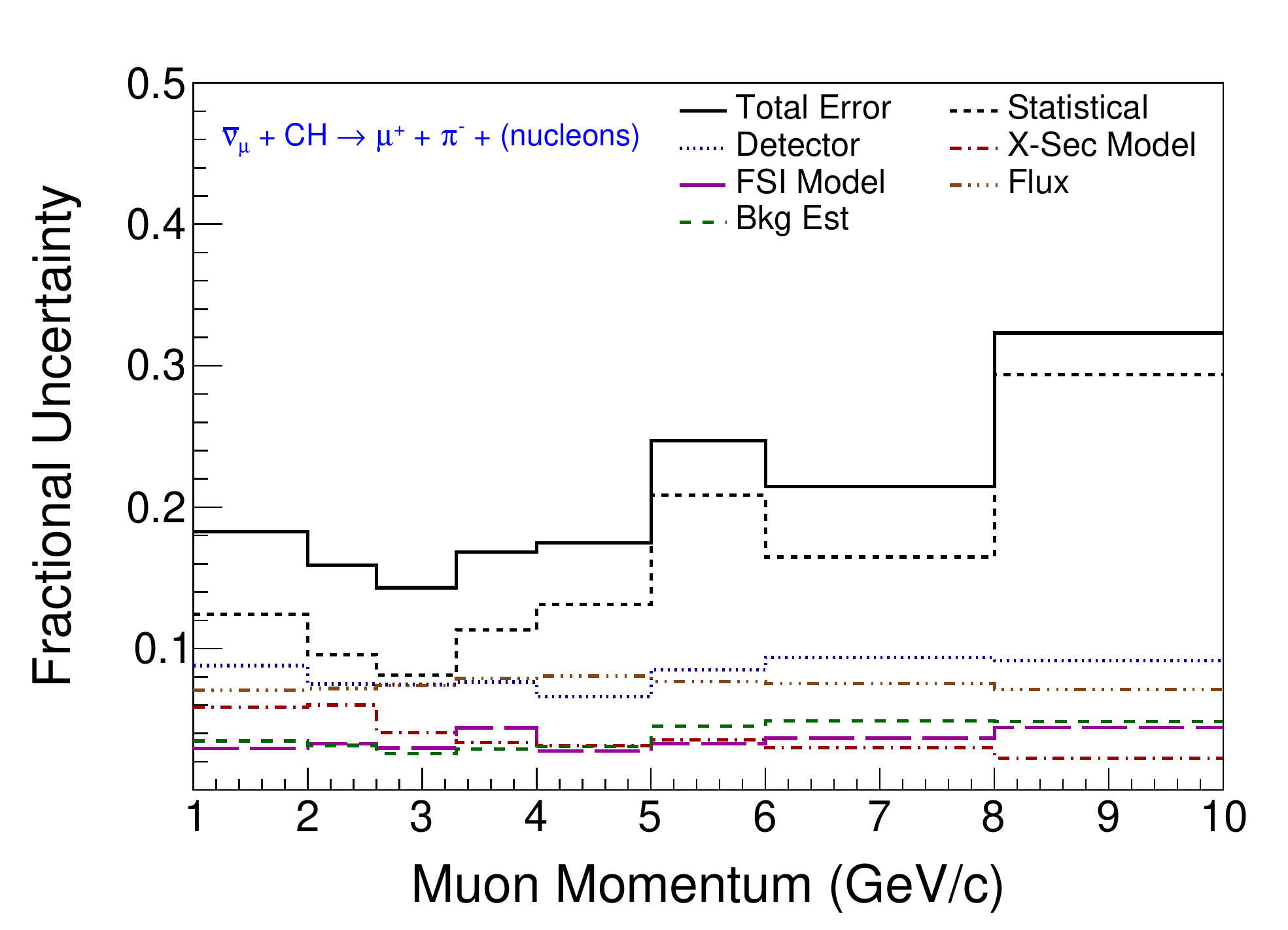} 
\includegraphics[width=8.5cm]{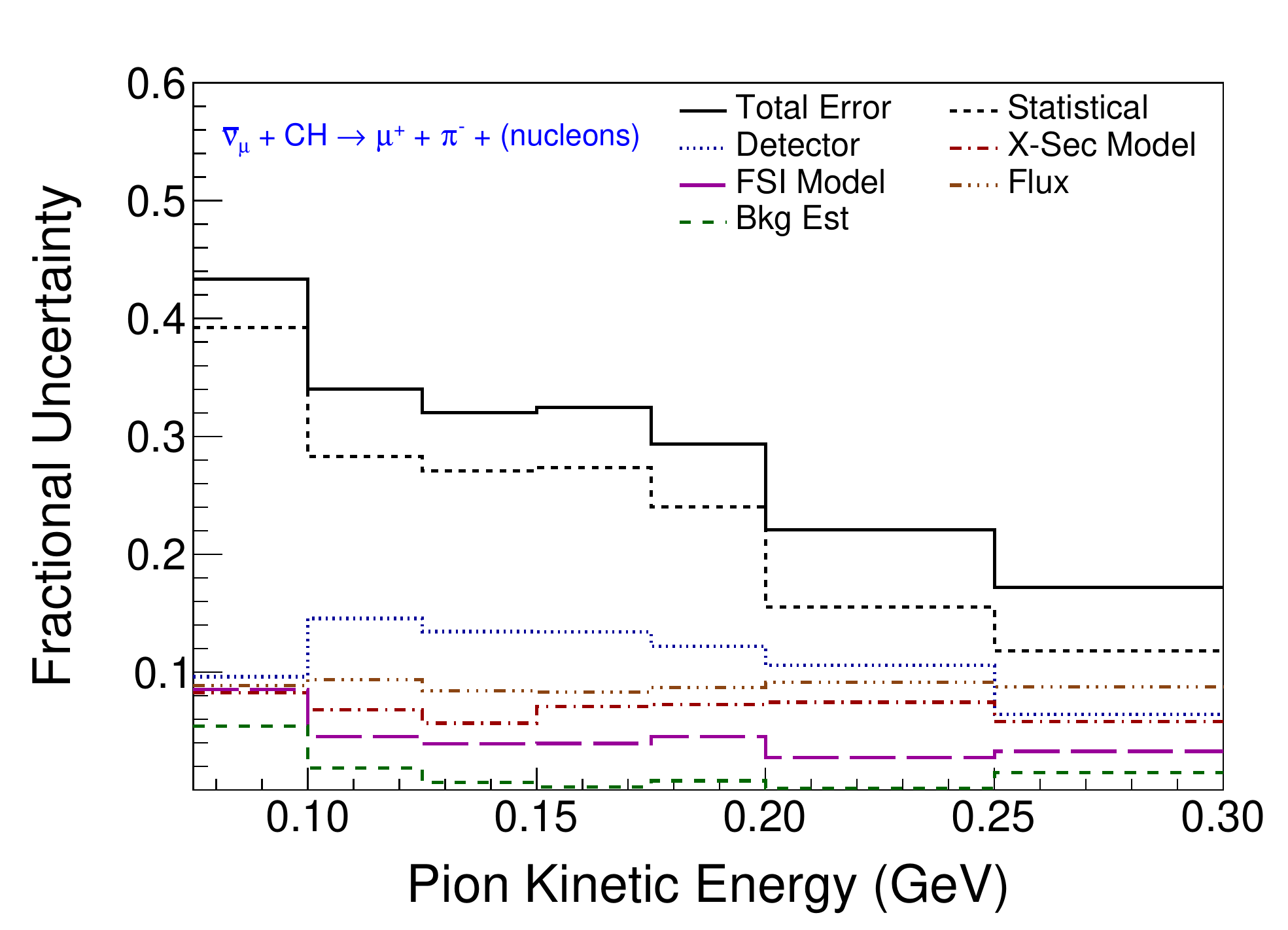}
\caption{Composition of fractional uncertainty in terms of systematic error categories plus the statistical uncertainty, 
for differential cross sections in $\mu^{+}$ momentum (upper plot) and $\pi^{-}$ kinetic energy (lower plot).
The statistical uncertainty (short-dash-line histogram) is the leading error source in all bins, with
detector response (fine-dash) and antineutrino flux (dot-dot-dash) uncertainties also contributing significantly.}
\label{Fig07}
\end{center}
\end{figure}

Uncertainty decompositions representative of cross-section determinations of directly measured kinematic variables 
are shown in Fig.~\ref{Fig07}, for $\mu^{+}$ momentum (upper plot) and for
charged pion kinetic energy (lower plot).  For all bins of either distribution, 
the finite data statistics (short-dash histogram) gives rise to larger uncertainties 
than does any single systematic category.   In particular, the large statistical error assigned to 
pion kinetic energies below 200 MeV reflects a large unfolding-correction uncertainty.
The detector response category contributes fractional uncertainties 
that range from 7\% to 9\% for muon momentum, and from 6\% to 15\% for pion kinetic energy.
Uncertainties assigned to the antineutrino flux are subject to constraints provided by the 
background normalization procedure.
Figure~\ref{Fig07} shows the fractional uncertainties from the flux 
and from the interaction cross-section model (GENIE) categories to be
constant or slowly varying over the measured ranges of $p_{\mu}$ 
and $T_{\pi}$, with value ranges of 7\% to 8\% and 8\% to $\leq$10\% respectively.

The differential cross sections of this work include $E_{\anu}$ and $Q^2$.   Since these variables are less
directly related to observations than are the muon and pion, their uncertainties have compositions that differ
somewhat from those shown in Fig.~\ref{Fig07}.   By way of illustration, the uncertainty decomposition 
for $E_{\anu}$ is shown in Fig.~\ref{Fig08}.   Here the statistical uncertainty dominates the low ($< 2.0$ GeV) and high ($> 6.0$ GeV)
neutrino energy bins, however in the $E_{\anu}$ range central to this work
the flux and detector response give fractional uncertainties of 9-12\% and 9\% respectably --
values that rival or exceed the statistical error.

\begin{figure}
  \begin{center}
\includegraphics[width=8.5cm]{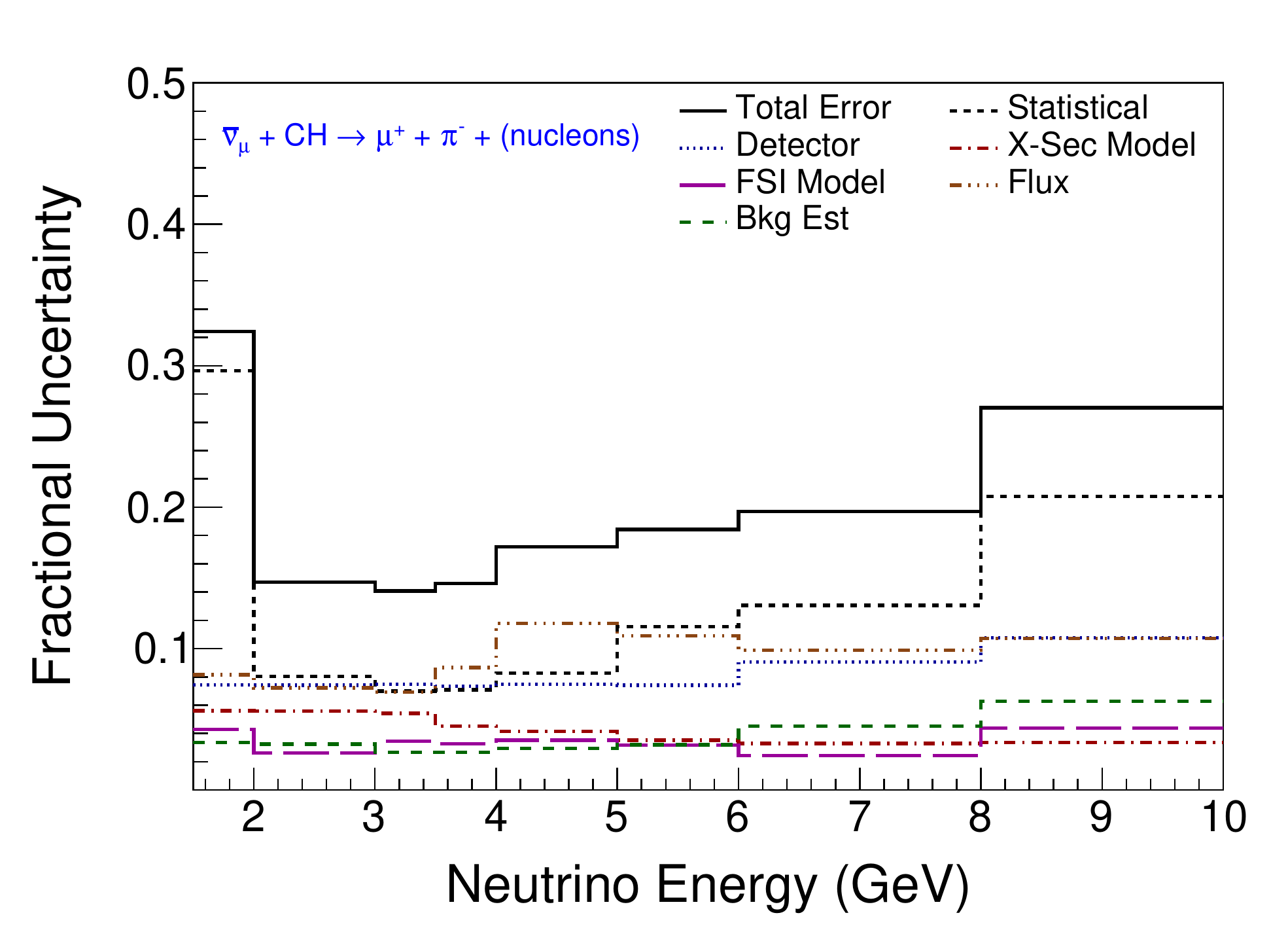} 
\caption{Bin-by-bin fractional uncertainty in systematic error categories plus statistical uncertainty, 
for cross section as a function of $E_{\anu}$.   The flux and detector response uncertainties 
are comparable to the statistical uncertainty in the 2.0 to 6.0 GeV range of $E_{\anu}$.}
\label{Fig08}
\end{center}
\end{figure}

The six uncertainty categories encompass all significant systematics of the analysis,
including the methodology by which nucleon kinetic energy is treated.   Nevertheless, it is of interest to quantify
the sensitivity of the $E_{\anu}$ determination to the reliance on kinematics for the inclusion of final-state
nucleon $T_{N}$.   For this purpose a simulation study was performed wherein
an uncertainty band for $T_{N}$ was assigned that covers the difference between binned values extracted by the analysis
versus MC true values.    Fractional uncertainties of 5\%, 10\%, and 25\% where allotted to $T_{N}$ ranges of 0-125 MeV, 
125-200 MeV, and $>$ 200 MeV respectively.    Simulation data for $T_{N}$ was then varied randomly in accord with the error
band and $E_{\anu}$ was recalculated.    The resulting r.m.s. spread in the fractional deviation of $E_{\anu}$ was less than 2.0\% overall, with 
deviations trending to higher values for $E_{\anu} >$ 5.5 GeV.   As Fig.~\ref{Fig08} clearly shows, an uncertainty of this magnitude is
well-covered by the ensemble of systematic and statistical uncertainties assigned to the $E_{\anu}$ measurement.

\begin{figure}
  \begin{center}
      \includegraphics[width=8.5cm]{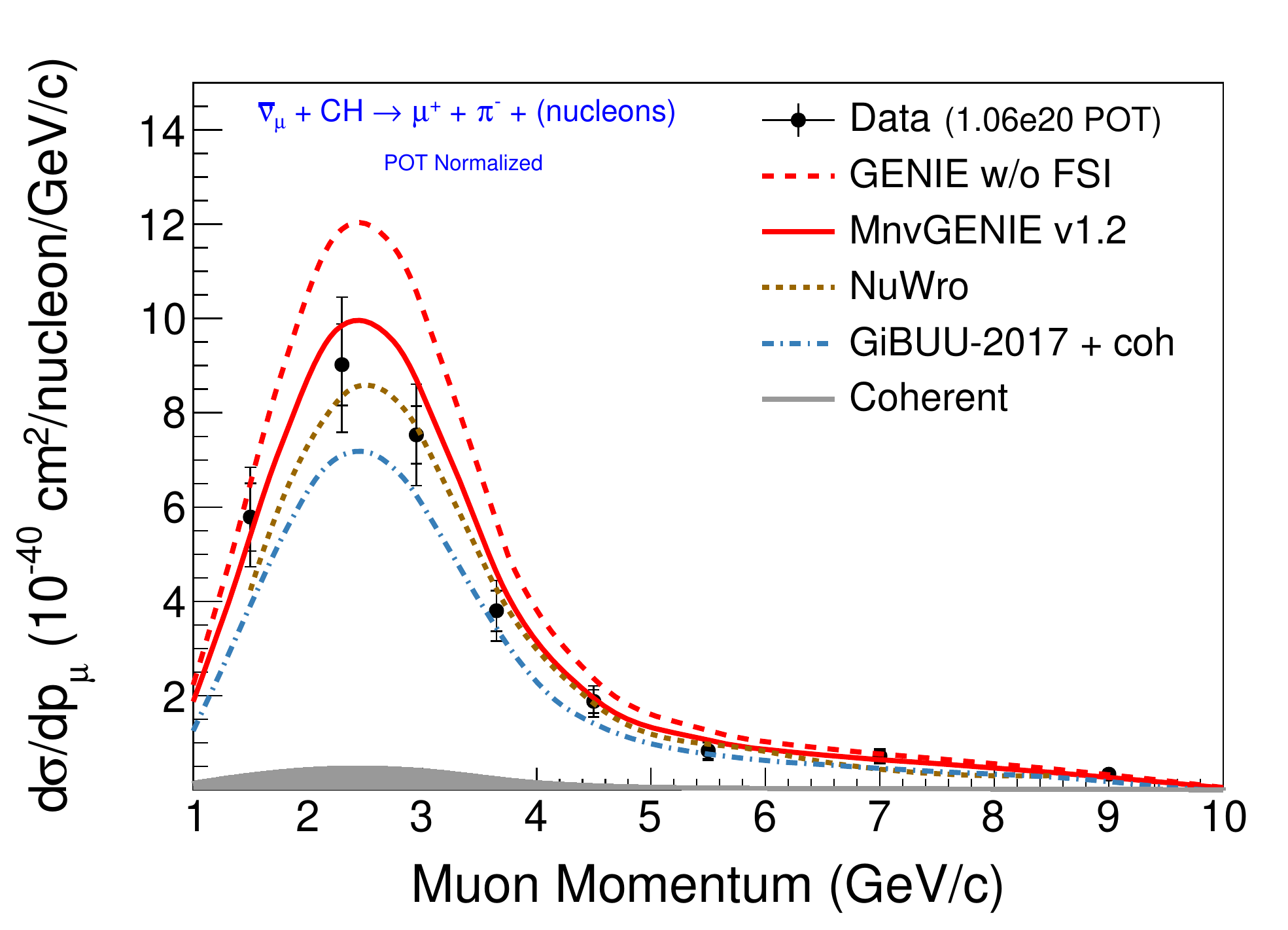}
    \caption{The flux-integrated muon-momentum differential cross section, $d\sigma/dp_{\mu}$
for muons with $\theta_{\mu} \leq 25^\circ$.  Data (solid circles) are shown with 
inner (outer) error bars that denote the statistical (total) uncertainties.         
The solid-line (dashed) curves show GENIE predictions with (without) FSI.   Short-dash and dot-dash curves show 
predictions by NuWro and GiBUU-2017.  The estimated contribution from CC coherent
scattering \eqref{coherent-pion-production} is given by the shaded region.}
    \label{Fig09}
  \end{center}
\end{figure}

\section{Muon Kinematics of $\anumu$-CC($\pi^{-}$)}
\label{sec:Muon-Kin}

\subsection{Muon momentum}
Figure~\ref{Fig09} shows the differential cross section for $\mu^{+}$ momentum, 
$d\sigma/dp_{\mu}$, of the signal channel.   The data are shown by the solid circles in the figure, with fully (partially) extended error bars
denoting the total (statistical) error associated with each data point.
Included in the cross section is a small event rate  
from CC coherent scattering reaction \eqref{coherent-pion-production} whose estimated contribution is 
indicated by the shaded area along the base of the distribution.
In accordance with the analysis signal definition, this differential cross section (and all others to follow)
is flux-integrated over the range 1.5\,GeV\,$\leq E_{\anu} \leq$\,10\,GeV, with the $\mu^{+}$ direction at production restricted 
to $\theta_{\mu} \leq 25^\circ$. 
The $\anumu$ flux spectrum strongly influences the shape of $d\sigma/dp_{\mu}$.   The distribution peaks 
near 2.5 GeV and then falls off rapidly as $p_\mu$ increases.   Predictions obtained with the GENIE-based MC are
shown by the two upper-most (red) curves in Fig.~\ref{Fig09}.   The dashed curve depicts a simulation in which 
pion and nucleon FSI effects are neglected.   It differs significantly from the full reference simulation with FSI included, shown
by the solid-line curve.    The difference is an average event-rate reduction of nearly $20\%$, reflecting
the strength of pion FSI in carbon, principally with $\pi^{-}$ absorption, for pions produced with kinetic energies in the region of
$\Delta(1232)$ excitation by $\pi^{-}$ intranuclear scattering.   With inclusion of FSI, the GENIE-based simulation still lies above the data,
giving an absolute event rate that exceeds the data by 8\%.
Allowing for the overestimate, one sees that the shape of the distribution is approximately reproduced for $p_{\mu} > 2$ GeV/c.

The short-dash and dot-dash curves in Fig.~\ref{Fig09} that lie below the GENIE prediction 
show expectations based on the NuWro and GiBUU-2017 event generators respectively.  NuWro does better than either GENIE 
or GiBUU-2017 with predicting the absolute data rate for most of the momentum range, with exception of momenta below 2 GeV/c where
GENIE matches the observed rate while the NuWro and GiBUU-2017 predictions fall below the data.    When each of the three generator predictions for
this differential cross section is area-normalized to the data (not shown), the generator curves nearly coincide and all three
generators give a good characterization of the distribution shape.

\begin{figure}
  \begin{center}
  \includegraphics[width=8.5cm]{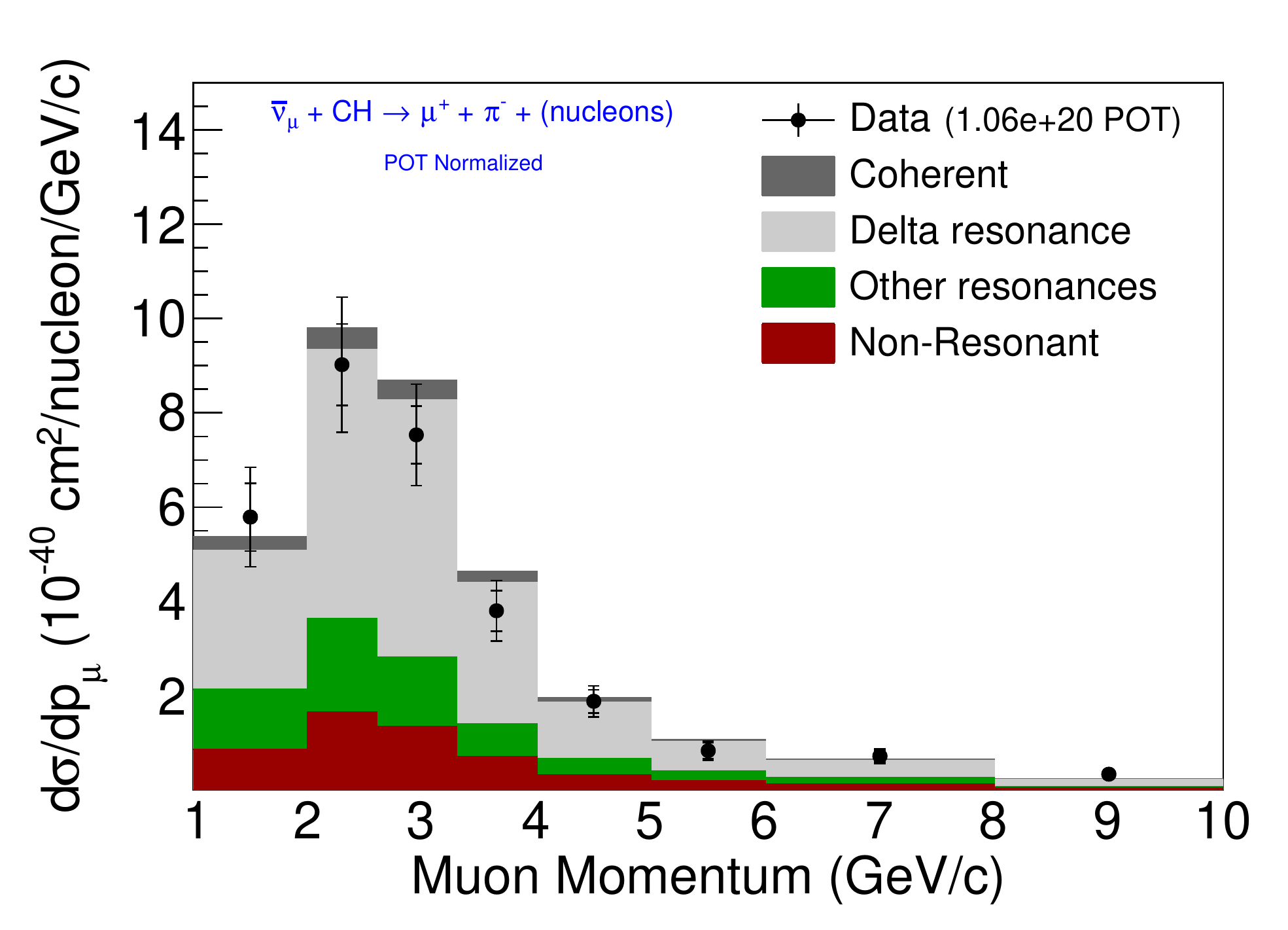}
   \caption{Cross section $d\sigma/dp_{\mu}$ as in Fig.~\ref{Fig09},  compared to component reaction processes of
   the reference simulation.   Production of $\Delta(1232)^{-}$ is predicted to dominate the signal 
     channel (gray-shade histogram) in all bins of muon momentum.}
        \label{Fig10}
  \end{center}
\end{figure}

The events of signal channel \eqref{signal-channel} can be characterized as originating from 
one of four processes:  {\it i)} pion production via the $\Delta(1232)$ resonance, 
{\it ii)} pion production via other baryon resonances, 
{\it iii)} Non-resonant pion production including DIS reactions, and 
{\it iv)} coherent pion production via reaction \eqref{coherent-pion-production}.  
Figure~\ref{Fig10} shows the relative strengths of these processes as predicted by the reference simulation.  
According to GENIE, $\Delta^{-}$ production accounts for  59\%
of the rate (upper, light-shade histogram in Fig.~\ref{Fig10});  production and decay of higher-mass $N^{*}$ resonances gives
an additional $\simeq 20\%$, with non-resonant pion production and CC coherent scattering accounting for the 
remaining 17\% and 4\% of the total rate, respectively.   These rates are for final states at emergence from target nuclei,
having been subjected to hadronic intranuclear scattering.   Their relationship to initially-produced final states is inferred
using the FSI model of the reference MC.    The relationship is well-illustrated by CC non-resonant single-$\pi^-$ events
wherein 12.5\%, 9.5\%, and 1.6\% portions of the initial sample migrate out of channel \eqref{signal-channel} as the result
of pion absorption, pion charge exchange, and of other hadronic FSI.

The four processes listed above are broadly distributed within the muon momentum distribution.
Figure~\ref{Fig10} indicates that the rate mis-match between GENIE and data could be alleviated by reducing
contribution(s) from the three non-coherent processes, but the data do not allow a unique prescription to be identified.

\subsection{Muon production angle}
\begin{figure}
  \begin{center}
    \includegraphics[width=8.5cm]{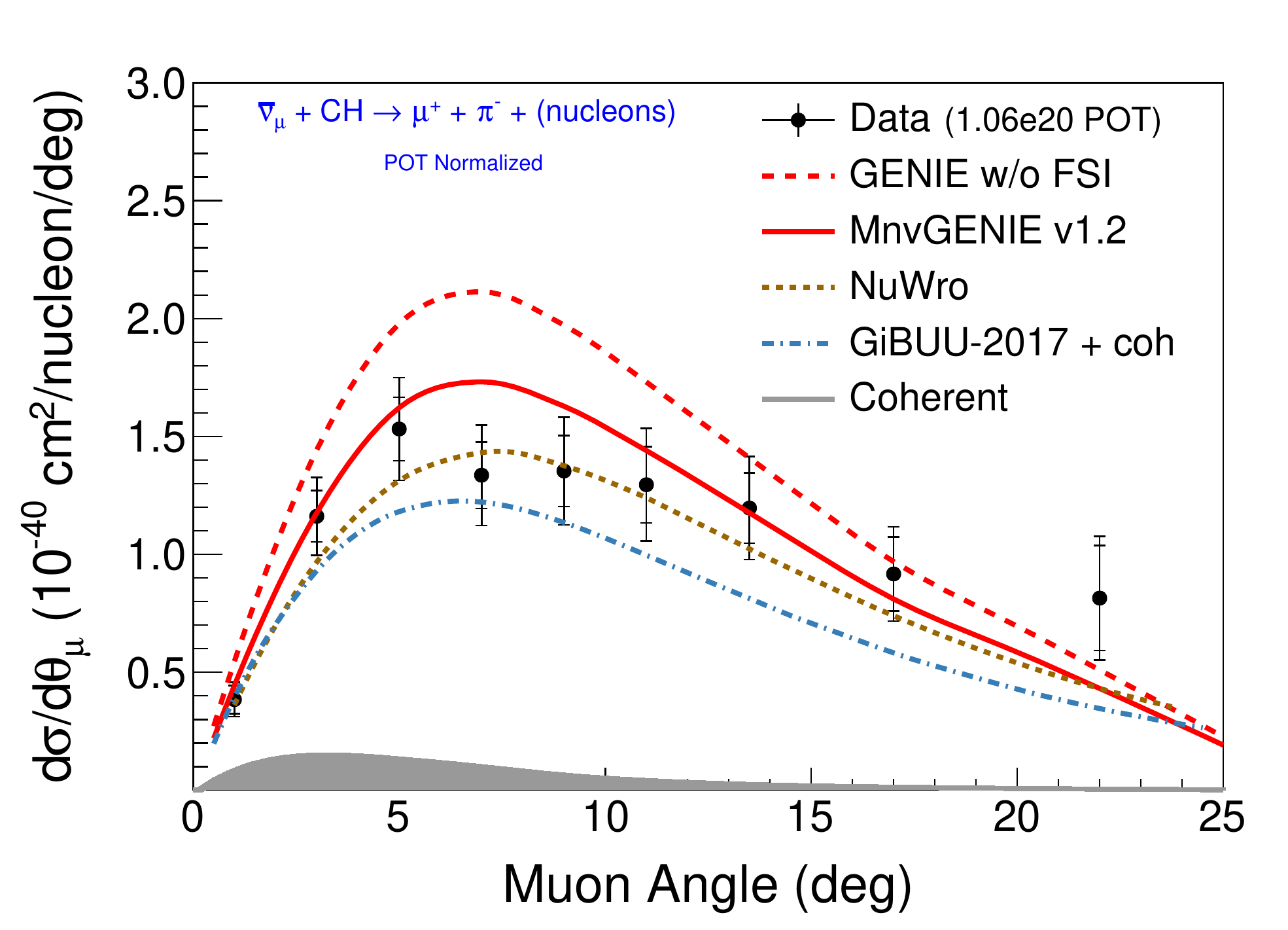}
    \caption{Differential cross section for muon production angle, $d\sigma/d\theta_{\mu}$.
    Data (solid circles) is compared to the predictions of GENIE with and without FSI (dashed, solid uppermost curves) 
    and with predictions from NuWro and GiBUU-2017.   The distribution shape is
    reproduced by all generators; NuWro comes closest with predicting the absolute event rate.}
    \label{Fig11}
  \end{center}
\end{figure}

Figure~\ref{Fig11} shows the $\mu^{+}$ differential cross section 
as a function of polar angle, $\theta_\mu$, with respect to the 
beam direction.  The distribution peaks near $7^\circ$ and then decreases gradually at larger angles.  

Comparison of GENIE, NuWro, and GiBUU-2017 predictions to the data show similar trends to those noted in Fig.~\ref{Fig09}.
All three generators give fairly accurate characterizations of the shape of $d\sigma/d\theta_{\mu}$, although the data above $\sim$\,$6^{\circ}$ 
exhibits a relatively flatter distribution.  Readily discernible is the over-prediction of absolute rate by GENIE and its under-prediction
by GiBUU-2017, with the closest agreement being 
achieved by NuWro.   The small contribution expected from CC coherent single-pion production (shaded region in
Fig.~\ref{Fig11}) is mostly confined to $\theta_\mu$ into forward angles $< 10^{\circ}$.
The fractional contributions from the three most prominent processes displayed in Fig.~\ref{Fig10} 
are predicted by GENIE to be nearly uniformly distributed over the measured angular range.   

The cross sections $d\sigma/dp_{\mu}$ and $d\sigma/d\theta_{\mu}$ can be compared to those previously reported by MINERvA
for $\anumu$-CC(1$\pi^{0}$) and for $\numu$-CC($\pi^{+}$) and $\numu$-CC(1$\pi^{0}$)~\cite{Carrie-pion, Altinok-2017}.
The observed spectral peaks roughly coincide for all four data sets, even though the
absolute cross sections are fairly different.   Differences in cross section magnitudes are certainly to be expected, since the four
pion production channels differ in their isospin compositions and in the role played by interferences between vector current
and axial vector current contributions, the latter being constructive in the $\numu$ channels and destructive in the $\anumu$ channels.

\section{Pion Kinematics of $\anumu$-CC($\pi^{-}$)}
\label{sec:Pion-Kin}

\begin{figure}
  \begin{center}
       \includegraphics[width=8.5cm]{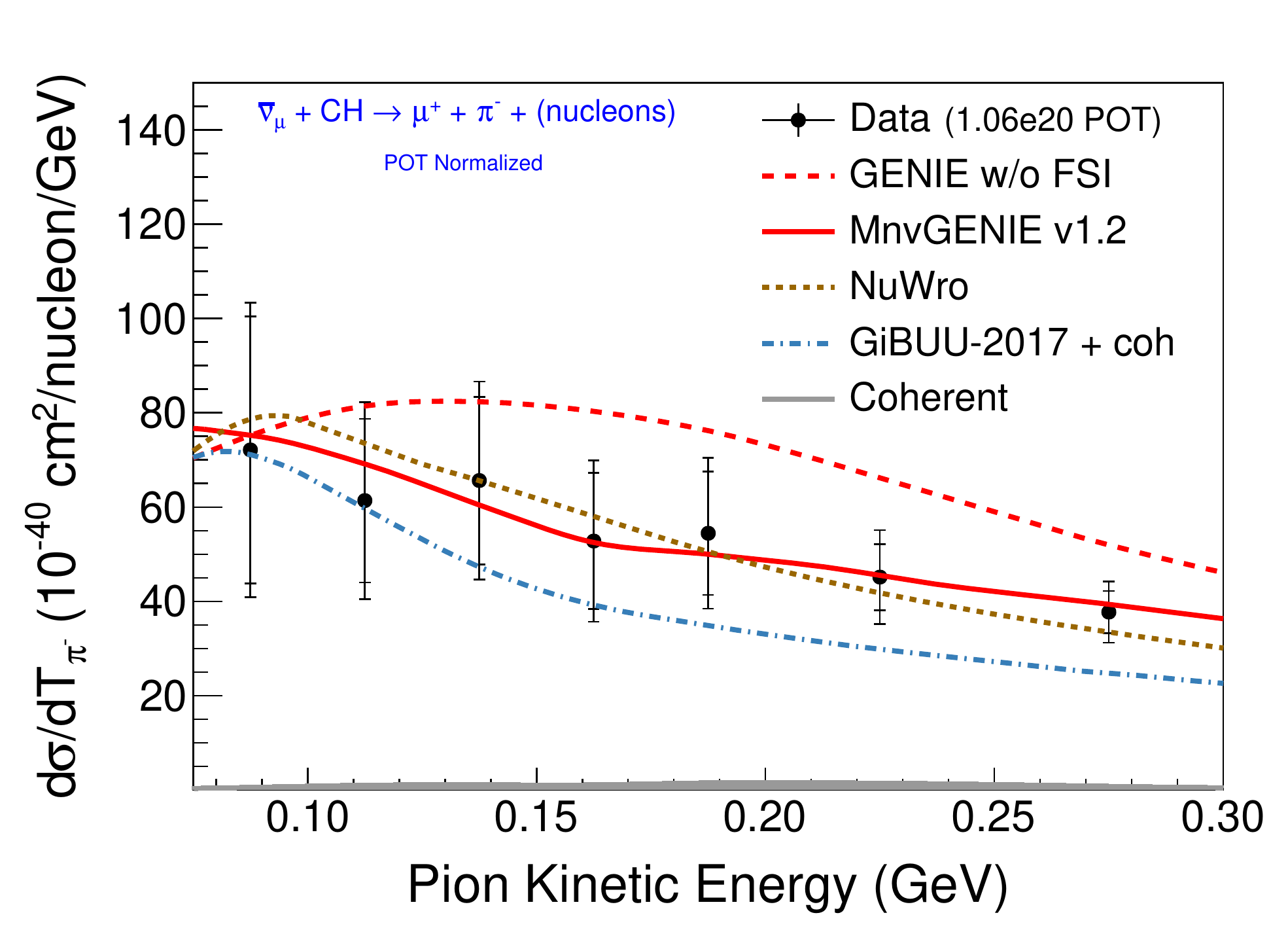}      
       \includegraphics[width=8.5cm]{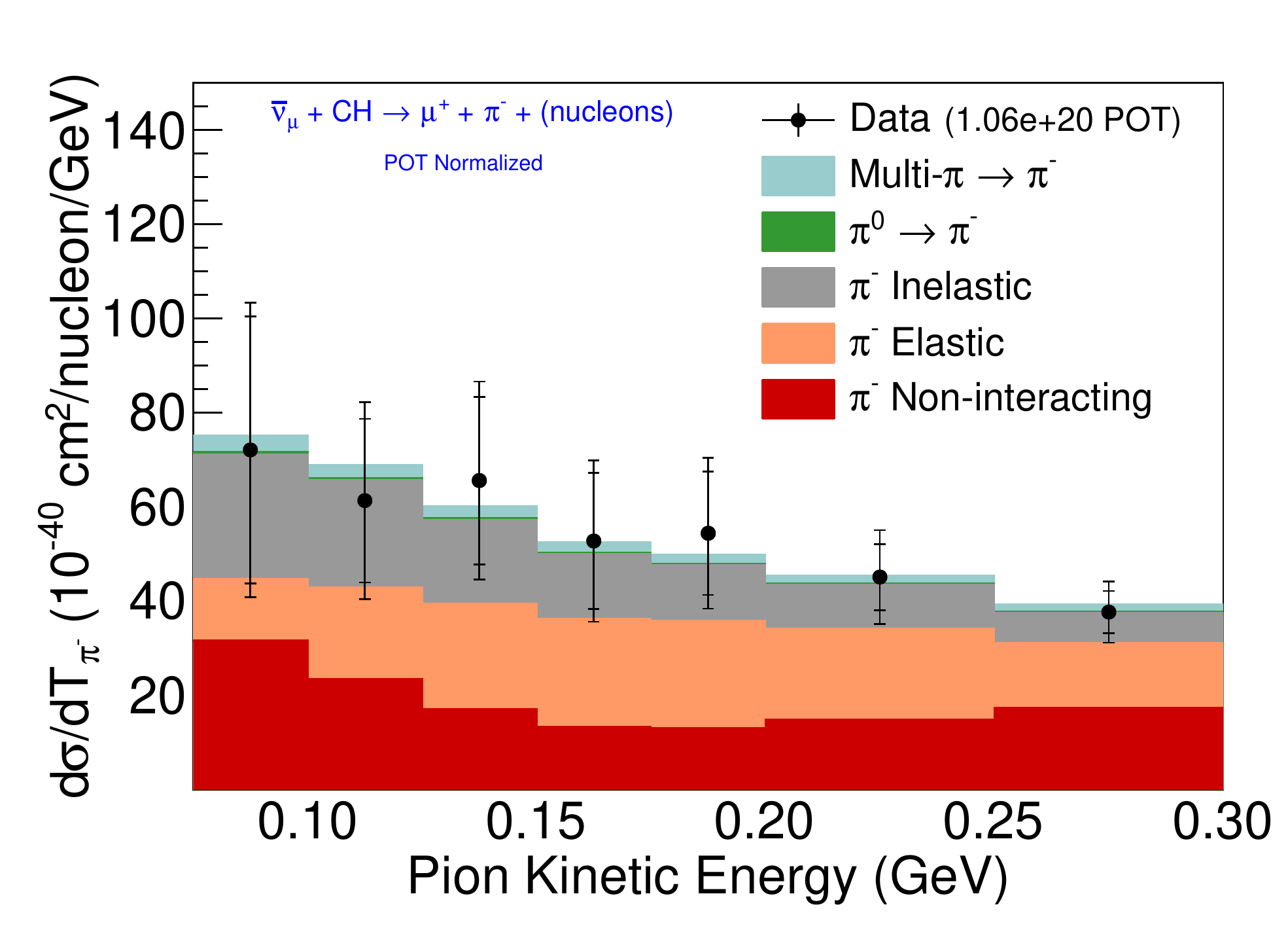}  
    \caption{Differential cross section $d\sigma/dT_{\pi^{-}}$ for pion kinetic energy.   Upper plot compares the data (solid points) 
     to predictions by the GENIE-based MC, NuWro, and GiBUU-2017.   Lower plot shows that GENIE achieves agreement 
     with measured $d\sigma/dT_{\pi^{-}}$ by combining pion FSI processes that differ in their component shapes.}
    \label{Fig12}
  \end{center}
\end{figure}

Figure~\ref{Fig12} shows the differential cross section for pion kinetic energy, $d\sigma/dT_{\pi^{-}}$.   Events in the lowest $T_{\pi^{-}}$ bin
have short $\pi^{-}$ tracks and their detection efficiency (2.8$\%$) is 2.4 times lower than that of the next higher bin.    The
efficiency correction to this bin mostly removes the depletion that appears in the initial data distribution for pion kinetic energy 
(lower-left plot of Fig.~\ref{Fig02}).   Additionally, the efficiency correction tends to flatten the remainder of the distribution.
The bin-by-bin uncertainties assigned to the data points are relatively large, reflecting the fact that the kinetic energy estimation
for $\pi^{-}$ tracks receives sizable corrections from the unfolding procedure.   The upper plot 
shows the gradually-falling shape of $d\sigma/dT_{\pi^{-}}$ to be reproduced by predictions from the generators, and the
absolute rate is roughly described.   The level of agreement provides support for the various FSI treatments for pions initiated within
carbon nuclei that are invoked by GENIE, NuWro, and GiBUU.

Produced $\pi^{-}$ mesons of the signal channel and the pions of background
reactions as well can undergo absorption, elastic and inelastic scattering, and/or charge exchange as they traverse the struck nucleus.
These pion FSI processes are especially prominent in range 90 MeV $< T_{\pi} < $ 210 MeV
corresponding excitation of the $\Delta$ in $\pi^{-}$ scattering on carbon~\cite{Binon-NP-1970}.
The agreement obtained by the GENIE-based MC for $d\sigma/dT_{\pi^{-}}$ is notable because the prediction represents a 
fairly intricate prediction that involves all pion subprocesses of the FSI model.

A breakdown of contributions from the component
processes is presented in the lower plot of Fig.~\ref{Fig12}.
The stacked histograms indicate that pions experiencing inelastic scattering, 
elastic scattering, or no scattering comprise the bulk of the sample (three lowest histograms), while
background feed-in from multiple-pion production with absorption
and from $\piz \rightarrow \pi^{-}$ charge exchange occurs with small rates (two uppermost histograms).
These processes are in addition to the significant amounts of absorption and charge-exchange that $\pi^{-}$ from initially produced signal events
are predicted to undergo.  According to the GENIE model, these latter processes have already winnowed down the signal sample from the initial interaction rate shown by the GENIE prediction without FSI (dashed curve in upper plot of Fig.~\ref{Fig12}), to give the rate predicted with FSI included -- depicted by the solid curve (upper plot) and the summed histograms (lower plot) of Fig.~\ref{Fig12}.   
Thus reproduction of the observed $\pi^{-}$ kinetic energy is achieved in the GENIE model by accounting for 
the combined effect of pion intranuclear elastic and inelastic scattering, charge exchange, absorption, together with instances of free pion
propagation through target carbon nuclei.

\begin{figure}
  \begin{center}
     \includegraphics[width=8.5cm]{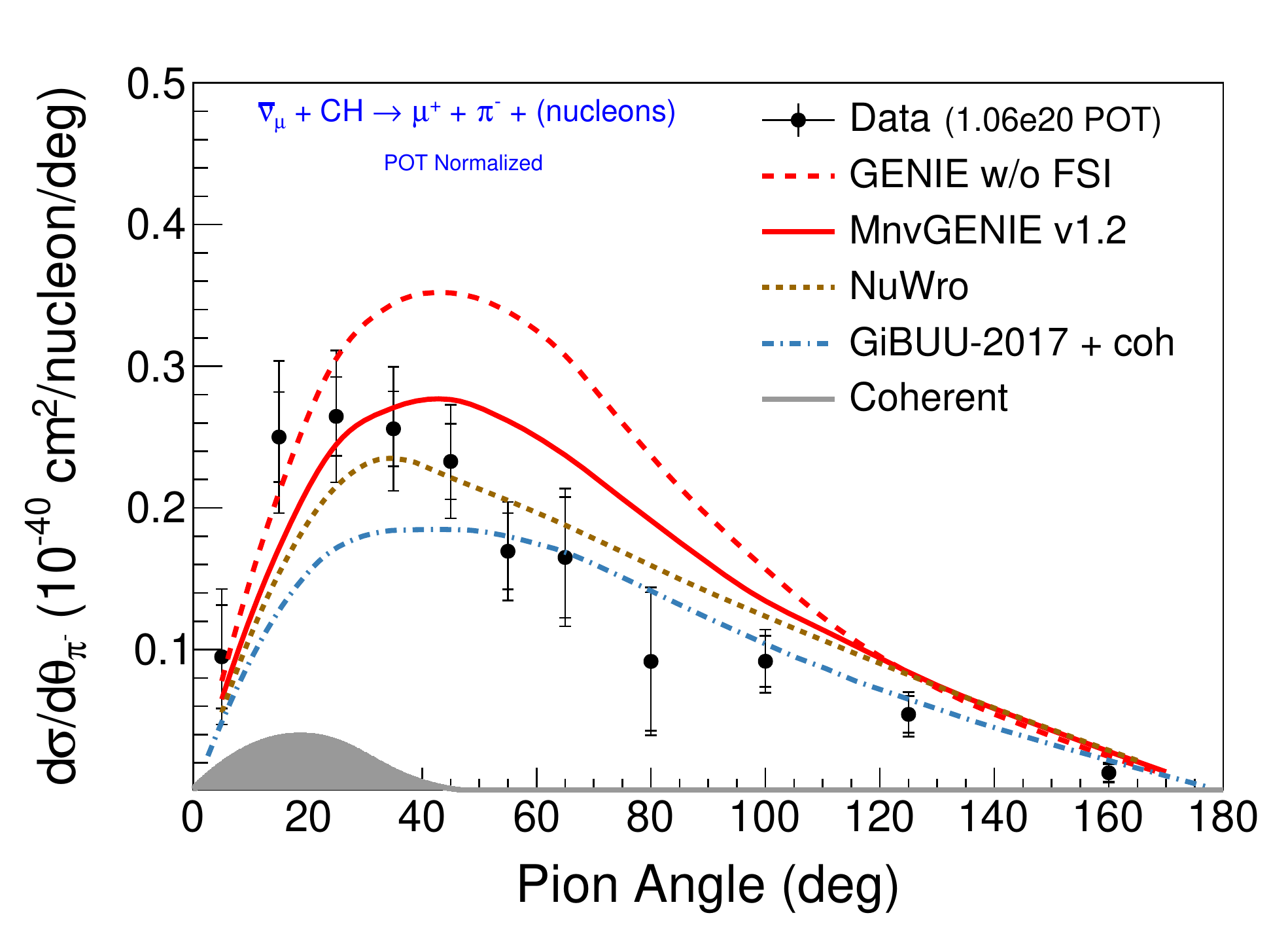} 
     \includegraphics[width=8.5cm]{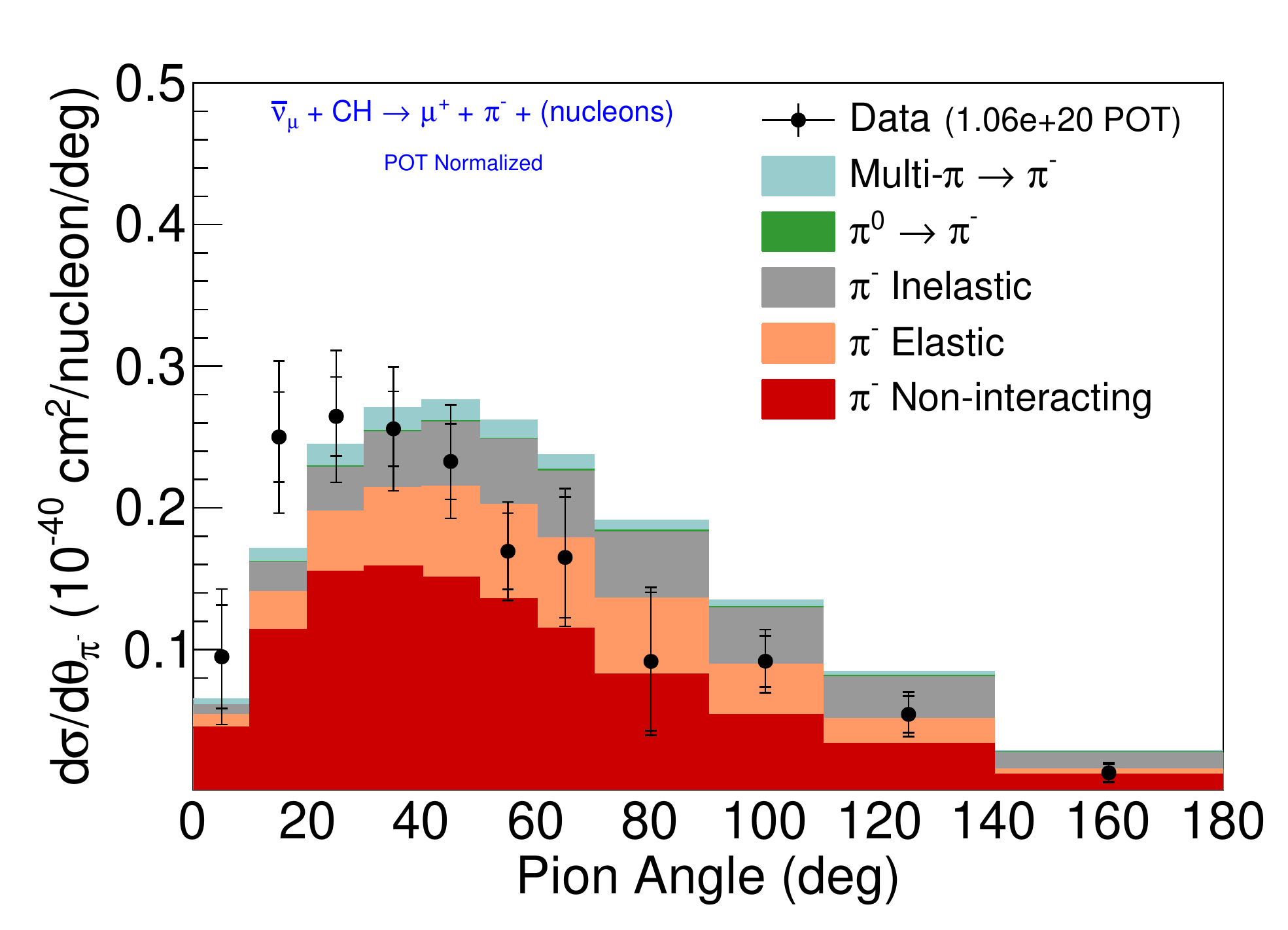}   
    \caption{Differential cross section for pion production angle.  Upper plot shows the data with 
    predictions from the GENIE-based MC and from NuWro and GiBUU-2017.  The gray-fill distribution
    depicts CC coherent scattering as measured by MINERvA.  Although coherent scattering is
    included in all the generator predictions, the data rate into forward $< 20^{\circ}$
    is underpredicted.     Lower plot shows contributions to $d\sigma/d\theta_{\pi^{-}}$ from component 
    pion FSI processes as estimated by the GENIE MC.  Coherent scattering is included in ``$\pi^{-}$ Non-interacting".}
    \label{Fig13}
  \end{center}
\end{figure}

Figure~\ref{Fig13} shows the differential cross section 
in pion angle measured relative to the $\anu$ beam direction.  
The data shows that most $\pi^{-}$s are produced in the forward hemisphere of the Lab frame, 
with angles around 30$^{\circ}$ being most probable.  The upper plot shows that the regions on either side of the
peak are not well-described by the event generators.   The data includes occurrences 
of CC coherent scattering via reaction \eqref{coherent-pion-production}, and this reaction is included in all of the generator
predictions displayed in the Figure.   
In particular, the CC coherent contribution measured by MINERvA is shown by the gray-fill distribution in the upper plot.
This contribution is included in the GENIE-based reference simulation shown by the solid curve in the upper plot.
It is also included as part of the ``$\pi^{-}$ Non-interacting" component displayed in the lower plot.
In the upper plot, the $\chi^{2}$ per degrees of freedom for the reference simulation with (without) FSI
is 24.2/11(47.8/11), while for NuWro and GiBUU-2017 it is 15.3/11 and 12.7/11 respectively.

The lower plot in Fig.~\ref{Fig13} decomposes the GENIE prediction into pion FSI processes, with ``pion non-interacting" (plus
coherently produced) being included as a process.    
None of the component processes are predicted to have angular features that change rapidly with increasing $\theta_{\pi^{-}}$.
Modeling of the inelastic and elastic FSI contributions include prescriptions for deflections of the initial pion direction.  Presumably these
could be adjusted to give a better description of the data.

\begin{figure}
  \begin{center}
      \includegraphics[width=8.5cm]{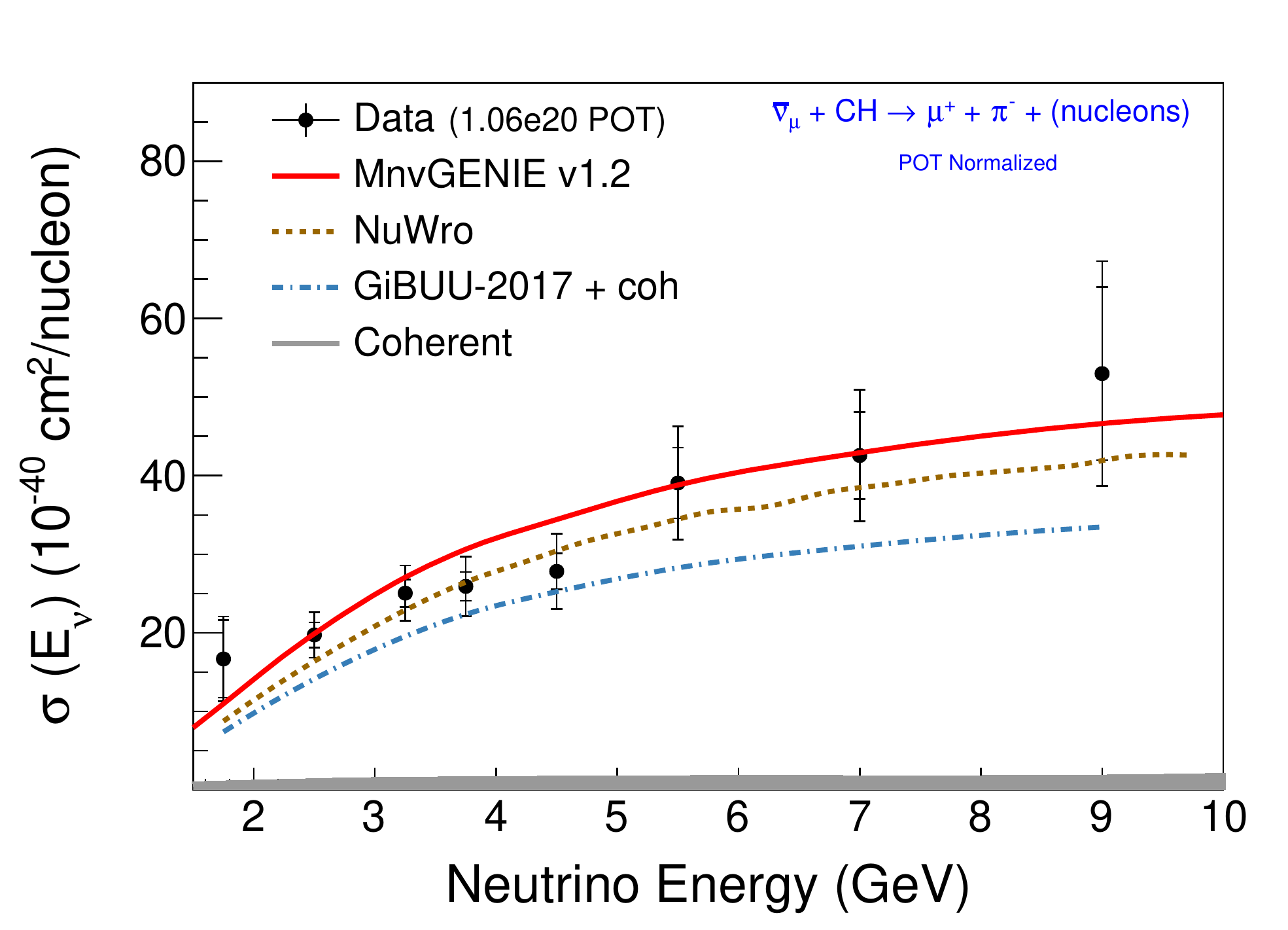}
        \includegraphics[width=8.5cm]{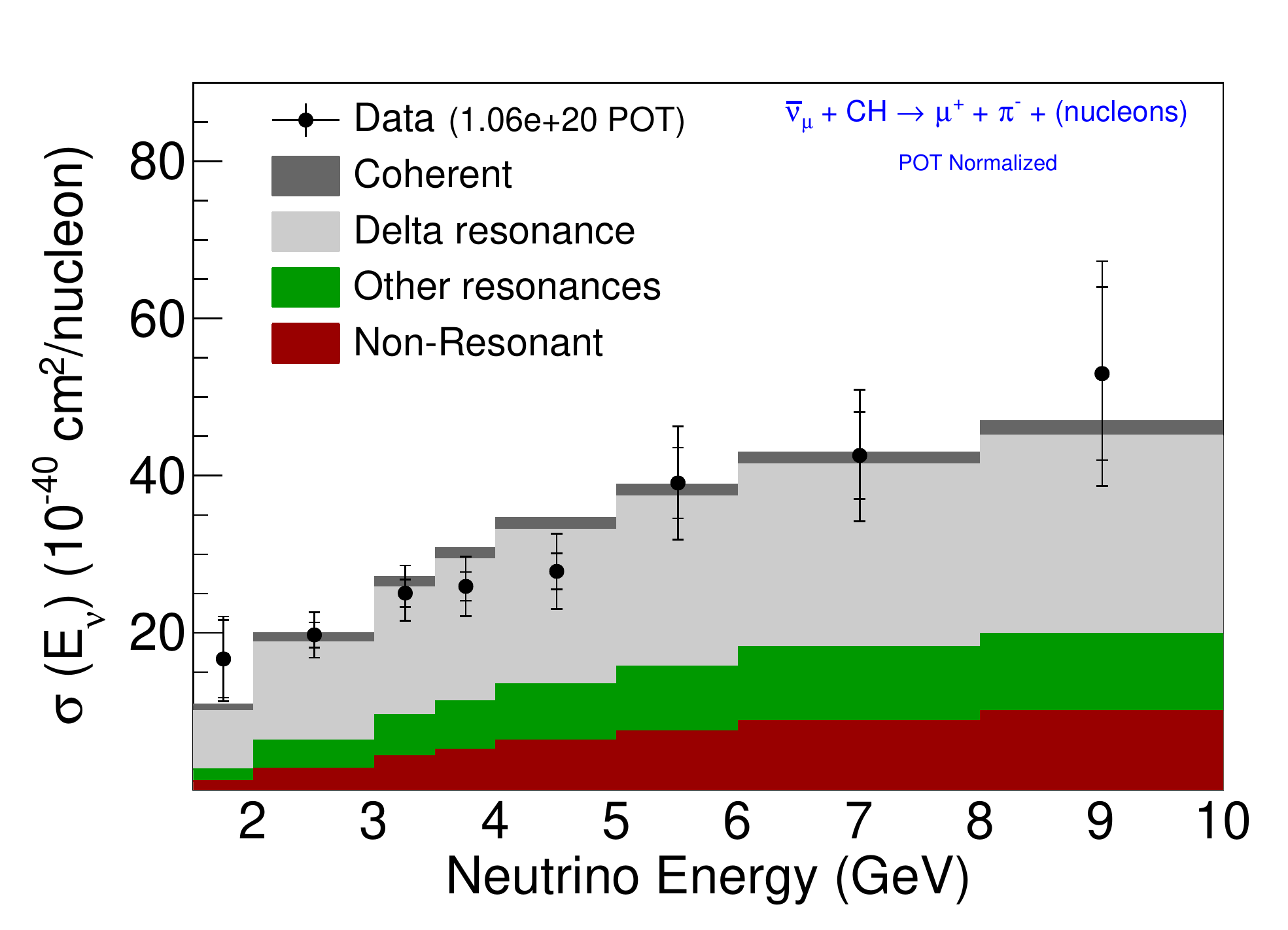}
    \caption{ Cross section (solid circles) as a function of antineutrino energy 
      for channel \eqref{signal-channel}.
       Upper plot compares the data to GENIE, NuWro, and GiBUU-2017 expectations.
       Lower plot shows contributions estimated by 
       GENIE from coherent scattering, $\Delta^{-}$ resonance production, N$^{*}$ states 
        above the $\Delta$, and pion non-resonance processes.}
           \label{Fig14}
  \end{center}
\end{figure}

\section{Cross sections for $E_{\anu}$ and $Q^2$}
\label{sec:Ev-Q2}

Figure~\ref{Fig14} shows the cross section as function of antineutrino energy, $\sigma(E_{\anu})$, for 
the signal sample, for which the invariant hadronic mass is restricted to $W_{exp} < 1.8\,$ GeV.
The data exhibit a gradual rise from threshold that continues with increasing $E_{\anu}$ to the end of the measured range at 10 GeV.
This behavior contrasts with the cross-section energy dependence of $\numu$-induced CC($\pi$) wherein the slope of $\sigma(E_{\nu})$ 
turns over and remains nearly zero above $\sim 5$ GeV~\cite{Carrie-pion, Altinok-2017}.    These differing trends reflect
the underlying vector minus axial vector $(V-A)$ structure of the hadronic current in $\Delta S = 0$ semileptonic interactions.  
The $VA$ interference terms contribute significantly to the cross sections at sub-GeV to few-GeV values of $E_{\anu}$, 
however they diminish rapidly relative to the $|V|^2$ and $|A|^2$ terms at higher incident (anti)neutrino energies.   
In contrast to $\numu$-induced CC($\pi$) cross sections, $VA$ interference terms
are of opposite sign and destructive for $\anumu$-CC($\pi$) interactions.   Consequently the slope turn-over point for cross sections of
antineutrino CC($\pi$) channels occurs at a distinctly higher incident energy than is observed with neutrino-induced CC($\pi$).

The three curves representing predictions based on GENIE, NuWro, and
GiBUU-2017 in Fig.~\ref{Fig14} (upper plot) exhibit the expected gradual rise of the cross section with $E_{\anu}$.
The GENIE-based reference MC is in agreement with the data with exception for the region between 3.5 to 5 GeV where 
offsets of order $1\,\sigma$ are indicated.   The NuWro prediction falls below the data in the two lowest $E_{\anu}$ bins, but matches the
data to within 1\,$\sigma$ throughout the higher $E_{\anu}$ range.   The GIBUU-2017 prediction, however, lies below the data at all energies.
The lower plot shows the relative cross-section portions that arise from the four interaction categories utilized by 
GENIE.    The relative contributions are predicted to remain in roughly constant proportion throughout the measured $E_{\anu}$ range, with 
$\Delta$ production being dominant throughout.

\begin{figure}
  \begin{center}
      \includegraphics[width=8.5cm]{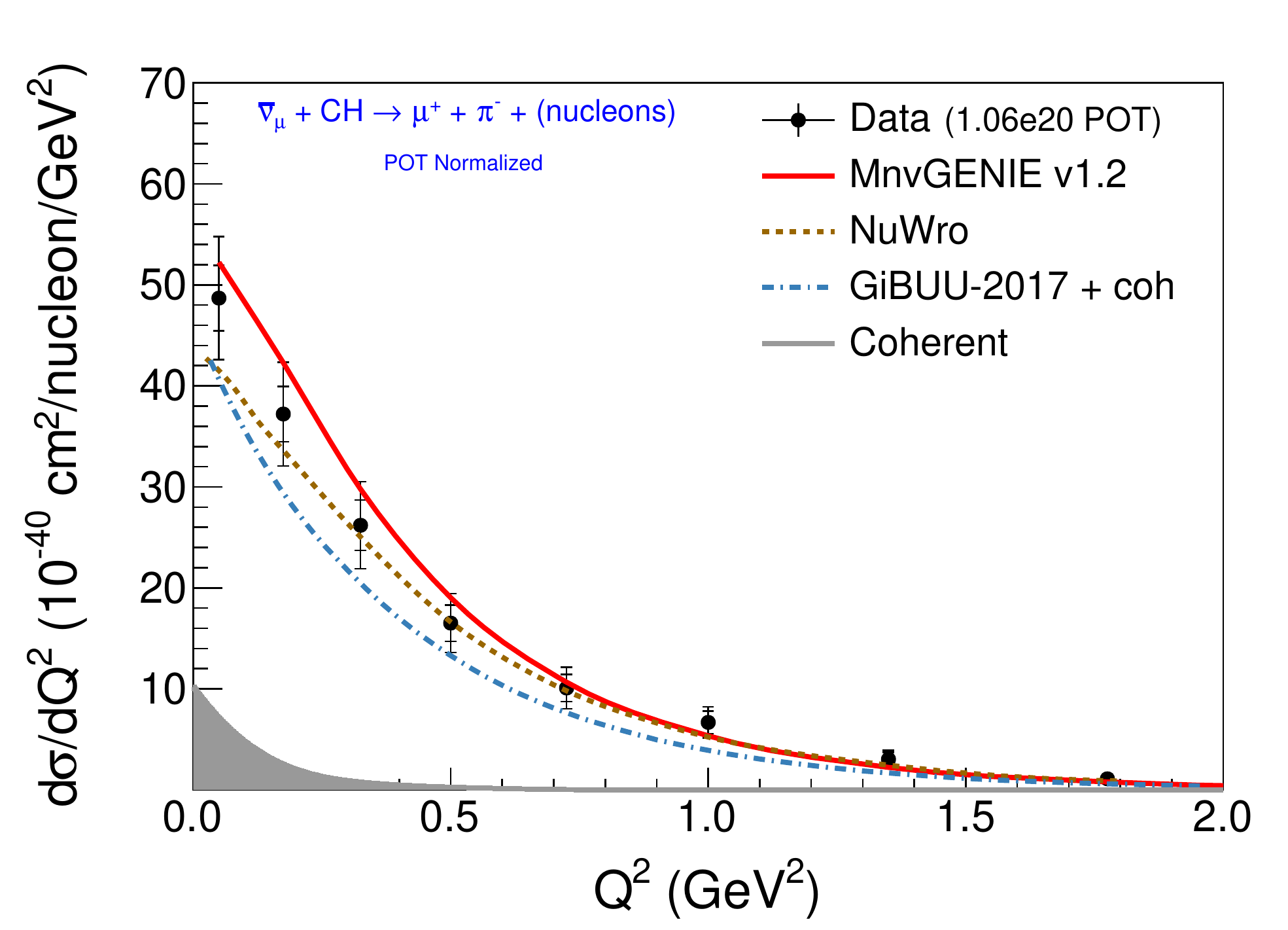}
            \includegraphics[width=8.5cm]{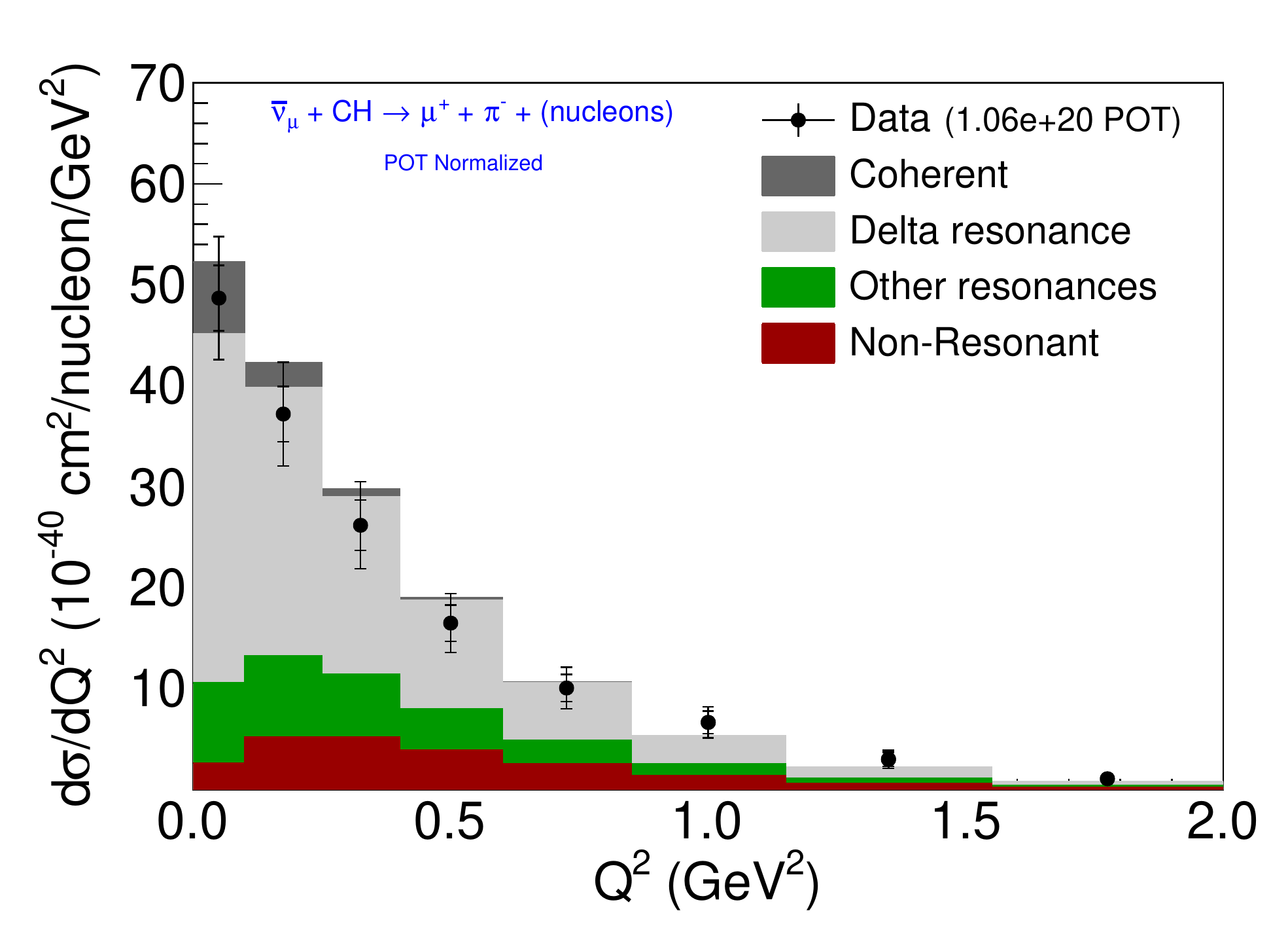}
    \caption{Differential cross section $d\sigma/dQ^{2}$ for the signal channel. Upper plot:  Predictions from the
    GENIE-based MC, NuWro, and GiBUU-2017 trend above, close to, and below the data respectively.
     Lower plot: Relative contributions from component processes according to GENIE.
     Coherent single-pion production is expected to contribute at very low $Q^{2}$.}                              
\label{Fig15}
\end{center}
\end{figure}

The squared four-momentum transfer from the lepton system, $Q^2$, 
is calculated using Eq.~\eqref{def-Q2};  
the differential cross section, $d\sigma/dQ^2$, is shown in Fig.~\ref{Fig15}.  
Comparisons with GENIE, NuWro, and GiBUU-2017 predictions 
are presented in the upper plot, and the relative contributions 
from the major reaction categories as estimated by GENIE 
are given in the lower plot.   A contribution from 
CC coherent scattering reaction~\eqref{coherent-pion-production} 
is estimated to occur in the region
$Q^{2} < 0.4\,$GeV$^2$.   The amount shown 
by the gray (dark gray) histograms in the upper (lower) plot is the rate
expected from MINERvA measurements~\cite{Mislivec-2018}.    
The data points in Fig.~\ref{Fig15} include this
CC coherent scattering contribution.

Even with allowance made for the presence of CC coherent scattering, the data 
do not exhibit a turn-over in $d\sigma/dQ^2$ as $Q^2$ approaches zero.  
The absence of a turn-over distinguishes the signal channel \eqref{signal-channel} of this work from the 
antineutrino and neutrino CC($\pi^{0}$) channels
previously studied by MINERvA~\cite{Carrie-pion, Altinok-2017}.   
This may be evidence for 
a process similar to CC coherent scattering that populates the
low $Q^2$ region of reactions \eqref{exclusive-channel-1} and \eqref{exclusive-channel-2}, 
but does not participate in reactions in which the target nucleon changes
its identity, such as $\anumu p \rightarrow \mu^+ \pi^0 n$.    
Charged-current diffractive scattering on nucleons is 
such a process, and its presence in high energy neutrino scattering 
has been pointed out by D. Rein~\cite{D-Rein-1986}.   
According to Rein, CC diffractive pion production 
must also be present in lower-$E_{\nu}$ scattering  
but its effect becomes very hard to disentangle from other CC($\pi$) processes.

In measurements of neutrino-induced CC$(\pi)$ channels carried out 
by MiniBooNE~\cite{MiniBooNE-piplus-2011, MiniBooNE-pi0-2011} and
by MINOS~\cite{ref:minos-QE}, it was found that MC agreement with data
can be improved by introducing, ad hoc, a suppression of baryon-resonance production at low $Q^2$.    
This approach finds some support from $Q^2$-dependent reductions that ensue with theoretical treatments of nuclear
medium effects that go beyond the 
Fermi gas model~\cite{BenharMeloni:2009,Nieves:2004wx, Martini:2009uj, Marteau:1999kt, ref:GandS-2003}. 
Figure~\ref{Fig15} suggests that low-$Q^2$ suppression may not be a universal feature of charged-current
pion production channels in $\numu/\anumu$ nucleus scattering.

\section{Estimation of $\anumu$-nucleon cross sections in hydrocarbon}
\label{sec:free-nucleon-scattering}

The definition of signal channel \eqref{signal-channel} that the analysis has used up to this point
refers to final-state topologies as they emerge from target nuclei.   
This signal definition is constructed such that all selections refer to directly observable quantities, and the differential cross sections 
subsequently presented refer to final-states that have been subjected to hadronic FSI.   Cross sections in this form provide
direct tests and feedback for continued development of neutrino event generators, 
as has been elaborated in Secs.~\ref{sec:Muon-Kin}, \ref{sec:Pion-Kin}, and \ref{sec:Ev-Q2}.

It is nevertheless of interest to investigate whether cross sections measured in a hydrocarbon medium 
can be related to the underlying initial antineutrino-nucleon interactions.   The CC($\pi$) cross sections reported by the bubble chamber
experiments of the 1970s and 80s, including those using propane-freon as well as deuterium or hydrogen fills, are entirely of the (anti)-neutrino
plus quasi-free nucleon kind~\cite{Gargamelle-1979, Pohl-nu-cc-pi-1979, SKAT-1989, 
Barish-antinu-1980, BEBC-D-antinu-1983, Allasia-1990, Allen-NP-1986}.   
Such measurements require fine-grained event imaging and rely upon certain aspects 
of neutrino-interaction modeling, e.g. Fermi motion and hadronic FSI.   Their pursuit has not been taken up
by spectrometer experiments of the modern era.   With the present analysis however, there arises motivation
to undertake determinations of the exclusive-channel cross sections for reactions \eqref{exclusive-channel-1} 
and \eqref{exclusive-channel-2}.   Two factors contribute to the feasibility of making these measurements with MINERvA:

\noindent
{\it (i)} Firstly, it is possible to relate the event rate determined 
for the signal channel into component rates for which the main contributors are 
the ``initial" (prior to FSI) quasi-free nucleon reactions \eqref{exclusive-channel-1} and \eqref{exclusive-channel-2}.    In this approach
the focus is placed on the initial $\anumu$-nucleon interactions that occur in target nuclei prior to any final-state alterations 
that may occur with the final-state hadrons as they traverse the parent nucleus.     These two initial reactions are now
to be regarded as ``the signal", while other initial reactions which, upon emergence from the parent nucleus, have morphed into 
channel \eqref{signal-channel}, are now regarded to be ``background".    The two aforementioned as-born signal reactions
differ according to the interaction nucleon that accompanies the muon and pion; the final-state hadronic
systems are (n $\pi^{-}$) and (p $\pi^{-}$) respectively.    Their different charge content gives a measurable
differences between distributions of vertex energy for the two final states.   While the distribution shapes must be 
taken from the reference simulation, the relative rates are well-constrained by fitting to the vertex energy
distribution observed in the signal sample, as is described in Sec.~\ref{sec:vertex-energy} below.

\smallskip
\noindent
{\it (ii)} Secondly, the GENIE-based reference MC appears to describe hadronic FSI in carbon rather well,
and the MC generally succeeds with shape predictions for backgrounds.  Importantly, there is no indication in
previous $\anumu$ and $\numu$ CC($\pi$) measurements of large spectral distortions arising from 2p2h production
~\cite{Trung-pion, Carrie-pion, Altinok-2017}.

These two factors are important because the analysis -- in order to ascertain the relative rates of 
the two initial, pre-FSI final states -- must rely on the hadronic FSI model of the reference simulation.

\smallskip

This approach is pursued in paragraphs below and cross sections are obtained 
for the exclusive reactions \eqref{exclusive-channel-1} and \eqref{exclusive-channel-2}.
Comparisons are made with measurements obtained with large bubble chambers. 

\smallskip

With exclusive-reaction cross sections for \eqref{exclusive-channel-1} and \eqref{exclusive-channel-2} in hand, it becomes possible
to relate them to the MINERvA measurement of $\anumu$-CC($\pi^{0}$) reported in Refs.~\cite{Trung-pion, Carrie-pion}.   Of course,
such a comparison requires the latter measurement to be subjected to the same approach --- one that elicits the underlying
initial reaction rate.    The opportunity then arises to decompose the three (non-coherent) exclusive reactions 
of $\anumu$-CC($\pi$) production in terms of the underlying isospin $I= 3/2$ and $I = 1/2$ amplitudes.     
A MINERvA-based isospin decomposition of $\anumu$-CC($\pi$) is
reported in Sec.~\ref{sec:Iso-Deco}.

\subsection{Channel separation using vertex energy}
\label{sec:vertex-energy}

The selected signal sample prior to background subtraction can be regarded as originating from four processes.
In addition to events of reactions \eqref{exclusive-channel-1} and \eqref{exclusive-channel-2}, there are contributions
from CC coherent scattering reaction \eqref{coherent-pion-production} and from background reactions.
The relative contributions of these processes to the signal channel rate can be distinguished by examining the 
``vertex energy" distribution of the signal sample.    For the purpose of this analysis, vertex energy is defined to be
the sum of energies of ionization hits deposited within 10\,cm of the primary vertex 
that is unassociated with the $\mu^{+}$ and $\pi^{-}$ tracks.   That vertex energy is a measurable quantity
is illustrated by the event displays in Fig.~\ref{Fig01}.

\begin{figure}
  \begin{center}
      \includegraphics[width=8.5cm]{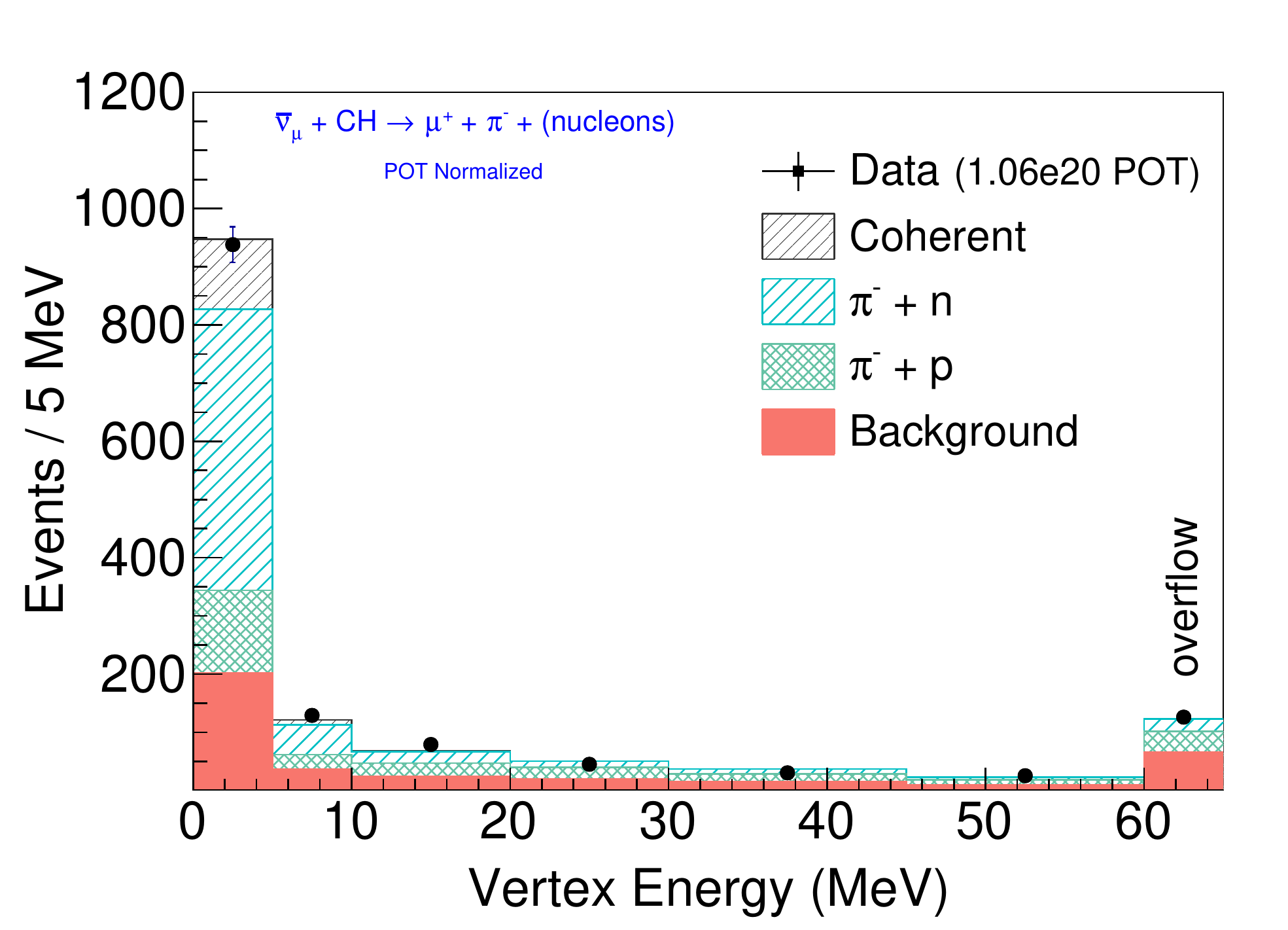}
            \includegraphics[width=8.5cm]{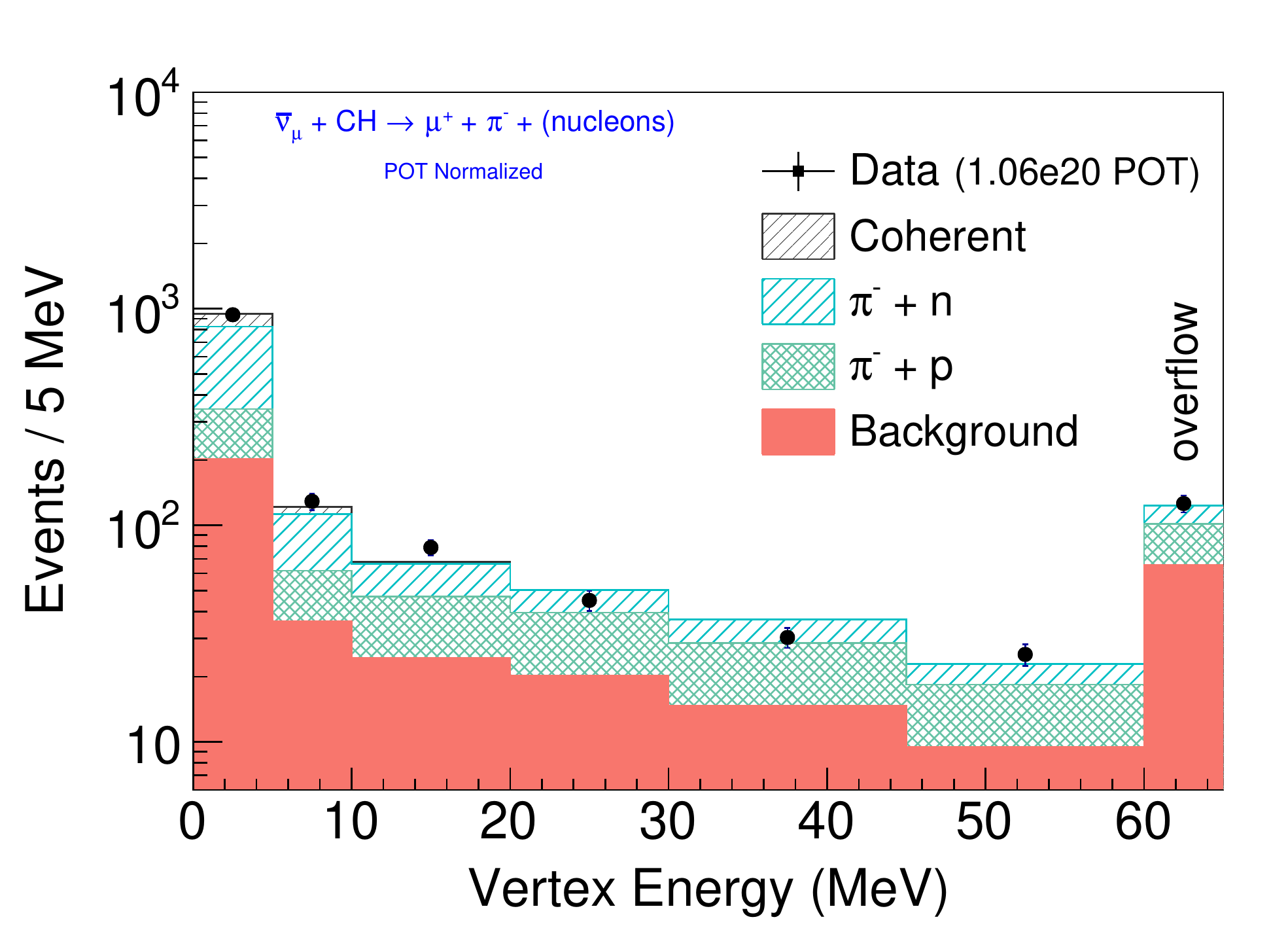}
 \caption{Distribution of event vertex energy in the signal sample (solid circles), displayed using linear and log scales 
 (upper, lower plots respectively).   Reference MC predictions for 
 contributions by reactions \eqref{exclusive-channel-1} and \eqref{exclusive-channel-2}, labeled
by their hadronic systems, are shown together with coherent 
scattering and background contributions.    The coherent contribution
is calculated from MINERvA measurement~\cite{Mislivec-2018}; 
the background rate is constrained by sideband fitting, and the exclusive-reaction rates 
are tuned to fit the signal sample data.}       
\label{Fig16}
\end{center}
\end{figure}

Figure~\ref{Fig16} shows the distribution of vertex energy in signal-sample candidates (solid circles, statistical errors).
In the upper plot, which displays the distribution using a linear scale, it is readily seen that
nearly two-thirds of the sample has $\leq 5$\,MeV of vertex energy and falls within the first bin. 
Events of the rest of the sample have vertex energies that lie in the higher range extending from 5 MeV to 
100 MeV.    In order to provide a clearer picture of this higher energy range, the same event distribution is 
displayed in the lower plot of Fig.~\ref{Fig16} using a logarithmic scale.

The MC component histograms in Fig.~\ref{Fig16} show the estimated contributions from the four processes.   
The breakout shown is obtained after three procedures have been applied:

\noindent
{\it (i)}   The coherent scattering contribution (top histogram, shaded) is fixed according to the measurement 
of reaction \eqref{coherent-pion-production} by MINERvA~\cite{Mislivec-2018}.  

\noindent
{\it (ii)}   The contribution from background is determined using 
a sideband constraint in the manner described for the main analysis, but with care taken concerning the signal definition
which for the present purpose has been changed.    Referring to the reference MC model for the sideband distribution
of vertex energy, the ``signal" are events that originated from reactions \eqref{exclusive-channel-1}, \eqref{exclusive-channel-2},
and \eqref{coherent-pion-production}, while everything else is background.   The distribution shapes for signal and background
are taken from the reference MC, and their absolute normalizations are determined by iterative fitting between
data of the sideband (to set the background normalization) and data of the analysis signal sample (to refine the estimate 
of signal contamination in the sideband).

\noindent
{\it (iii)}  With the background and coherent scattering contributions thereby set, a fit to the vertex energy data is performed 
wherein the distribution shapes for reaction \eqref{exclusive-channel-1} and \eqref{exclusive-channel-2} contributions
are taken from the reference simulation, and their normalizations are used as fit parameters.          

It is readily seen in Fig.~\ref{Fig16} that the fit adjustment of the MC model gives a good description of the data.
Based on this description, the numbers of interactions \eqref{exclusive-channel-1} and \eqref{exclusive-channel-2} 
that underwrite the signal-sample population are estimated to be 
$N(\mu^{+} n \pi^{-}) = 682  \pm 121$ and 
$N(\mu^{+} p \pi^{-}) = 349 \pm 121$, where the error bars include systematic as well as statistical uncertainties.
To convert these event counts into cross sections, it is required to know the efficiencies with which
the analysis selection chain retains the progeny of reactions \eqref{exclusive-channel-1} and \eqref{exclusive-channel-2} 
and allows them to appear in the selected signal sample.    These efficiencies, as estimated by the reference simulation, 
are $\epsilon(\mu^{+}\pi^{-}\textrm{n}) = 4.9\%$ and $\epsilon(\mu^{+}\pi^{-}\textrm{p}) = 4.1\%$.
The hydrocarbon target region of MINERvA contains 15\% more protons than neutrons.   The difference is taken into 
account in order to obtain exclusive-channel cross sections that are ``per nucleon" for an isoscalar target medium.
The cross-section values are:
\begin{equation}
\label{xsec-exclusive-channel-1}
\sigma(\mu^{+}\pi^{-}\textrm{n}) =  19.7 \pm 4.4 \times10^{-40}\,\text{cm}^2~\text{per nucleon},
\end{equation}
\begin{equation}
\label{xsec-exclusive-channel-2}
\sigma(\mu^{+}\pi^{-}\textrm{p}) =  12.1 \pm 4.5 \times10^{-40}\,\text{cm}^2~\text{per nucleon}.
\end{equation}
Comparable results are the
flux-averaged cross sections for $W < 2$\,GeV based on Gargamelle antineutrino data.   These are stated without errors
in Table VII of Ref.~\cite{Rein-Sehgal-1981} as follows: $\sigma(\mu^{+}\pi^{-}\textrm{n}) =  25.1 \times10^{-40}\,\text{cm}^2$ and 
$\sigma(\mu^{+}\pi^{-}\textrm{p}) =  10.1 \times10^{-40}\,\text{cm}^2$.   Table 3 and Figs. 2 and 3 of Ref.~\cite{Gargamelle-1979},
indicate uncertainties for these cross sections (arising from background correction, nuclear effects, and finite statistics) to be of order 25\%.

\section{Isospin composition of $\anumu$-CC($\pi$)}
\label{sec:Iso-Deco}

A broader perspective on $\anumu$-CC($\pi$) reactions can be obtained 
by relating the MINERvA measurement of $\anumu$-CC($\pi^{0}$)~\cite{Trung-pion, Carrie-pion}
to cross sections \eqref{xsec-exclusive-channel-1}  and \eqref{xsec-exclusive-channel-2}.
To this end, a reanalysis of the latter data has been carried out to extract the free-proton target cross section for
the exclusive channel
\begin{equation}
\label{exclusive-channel-3}
\anumu + \textrm{p} \rightarrow\mu^{+}+\pi^{0}+\textrm{n} .
\end{equation}
The measured signal channel of $\anumu$-CC($\pi^{0}$) is devoid of any coherent scattering contribution, and 
exclusive reaction \eqref{exclusive-channel-3} is the only $\anumu$-nucleon interaction that feeds the
signal channel.     Consequently the extraction of the reaction \eqref{exclusive-channel-3} cross section is 
relatively straightforward.  The event selections described in 
Secs.~\ref{sec:Reco-and-select-1}, \ref{sec:Reco-and-select-2}, and \ref{sec:Estimate-E-Q-W-signal-sample} are applied in the same way
to the data of the earlier work. 
   As previously noted, a weight is applied to normalize the cross section
for reaction \eqref{exclusive-channel-3} to describe scattering per nucleon from an isoscalar target.
The `as born' free-nucleon target cross section for
reaction \eqref{exclusive-channel-3} thereby obtained is
\begin{equation}
\label{xsec-exclusive-channel-3}
\sigma(\mu^{+}\pi^{0}\textrm{n}) =  10.7 \pm 1.7 \times10^{-40}\,\text{cm}^2~\text{per nucleon}.
\end{equation}
The flux-averaged value for $W < 2$\,GeV attributed to Gargamelle~\cite{Rein-Sehgal-1981}  is
$\sigma(\mu^{+}\pi^{0}\textrm{n}) =  9.5 \times10^{-40}\,\text{cm}^2$.

The cross sections \eqref{xsec-exclusive-channel-1}, \eqref{xsec-exclusive-channel-2}, 
and \eqref{xsec-exclusive-channel-3} as 
hereby extracted from MINERvA data, 
comprise the complete set of free-nucleon cross sections for exclusive $\anumu$-CC($\pi$) reactions.
Each of these reactions proceeds via the $\Delta S = 0$ weak hadronic charged current;  The current operator transforms
as an isovector.   This has the consequence that the final states of  \eqref{exclusive-channel-1}, 
\eqref{exclusive-channel-2}, and  \eqref{exclusive-channel-3} 
can be expressed in terms of reduced amplitudes $A_{3}$ and $A_{1}$
which describe the $I = 3/2$ and $I = 1/2$ states of the $\pi N$ system.   These amplitudes 
(in the convention of Rein-Sehgal~\cite{Rein-Sehgal-1981}) can be written as
\[ \begin{split}
&A(\anu n\rightarrow \mu^{+}n\pi^{-}) = \sqrt{2} A_{3}, \\
&A(\anu p\rightarrow \mu^{+}n\,\pi^{0}) ~= \frac{2}{3} (A_{3} - A_{1} ), \\
&A(\anu p\rightarrow \mu^{+}p\pi^{-}) = \frac{\sqrt{2}}{3}(A_{3} + 2 A_{1}).
 \end{split} \]
Relations are thereby implied that interrelate these cross sections.    For example,
if the $\Delta(1232)$ dominates a selected kinematic region such that $| A_{3}| >> |A_{1}|$,
then one expects certain cross-section ratios to exhibit particular values.   Specifically, for the ratios
\[ \begin{split}
&R_{1} \equiv \sigma (\mu^{+}n\pi^{0})/ \sigma (\mu^{+}n\pi^{-}), ~~\text{and}~~\\
 &R_{2} \equiv \sigma (\mu^{+}p\pi^{-})/ \sigma (\mu^{+}n\pi^{-}),
 \end{split} \]
one expects $R_{1} \simeq 2/9$ and $R_{2} \simeq 1/9$ for the case of $A_{3}$ dominance.
As shown below, the data does not support this particular scenario.

More generally, the flux-averaged free-nucleon cross sections 
for \eqref{exclusive-channel-1}, \eqref{exclusive-channel-2}, and \eqref{exclusive-channel-3} 
in the hadronic mass range $W < 1.8$\,GeV, enable values to be obtained 
for the following averaged quantities~\cite{Rein-Sehgal-1981}:
\[ \begin{split}
&\braket{ |A_{3}|^2} \, = \frac{1}{2} \sigma(\mu^{+}n\pi^{-}), \\
&\braket{ |A_{1}|^2} \, =  \frac{3}{4} \big\{ \sigma(\mu^{+}n\pi^{0})  +  \sigma(\mu^{+}p\pi^{-}) -  \frac{1}{3} \sigma(\mu^{+}n\pi^{-}) \big\}, \\
&\braket{ \mathcal{R}e(A_{3}^{*} A_{1})} \,= \\
&~~~~~~~~~~~~\frac{3}{8} \big\{ \sigma(\mu^{+}p\pi^{-}) - 2\, \sigma(\mu^{+}n\pi^{0}) + \frac{1}{3} \sigma(\mu^{+}n\pi^{-}) \big\}.
\end{split} \]
The relative magnitude of the two isospin amplitudes, $R^{\anu}$, and their relative phase, $\phi^{\anu}$, are given by the relations
\[ \begin{split}
&R^{\anu} =  \big\{\braket{|A_{1}|^2} / \braket{|A_{3}|^2}\big\}^{1/2}, \\
&\cos \phi^{\anu} =\,\, \braket{\mathcal{R}e(A_{3}^{*} A_{1})} / \braket{|A_{3}|^2}^{1/2} \braket{|A_{1}|^2}^{1/2}.
\end{split} \]
The above quantities can be written as functions of the $CC(\pi)$ cross sections or
as functions of $R_{1}$ and $R_{2}$.   (See Eqs.~(4.8) and (4.9) of Ref.~\cite{Rein-Sehgal-1981}.)

\subsection{MINERvA results}
\label{subsec:M-results}

Using the cross-section values \eqref{xsec-exclusive-channel-1}, \eqref{xsec-exclusive-channel-2}, and \eqref{xsec-exclusive-channel-3} 
this analysis obtains $R_{1} = 0.46 \pm 0.08$ and $R_{2} = 0.52 \pm 0.19$. 
The relative magnitude and phase of the isospin amplitudes are then determined to be
\begin{equation}
\label{magnitude-phase-1}
R^{\anu} = 0.99 \pm 0.19, ~~~ \phi^{\anu} = 93^{\circ} \pm 7^{\circ}.
\end{equation}

The $R^{\anu}$ value indicates a large presence for the $I = 1/2$ amplitude in the final states
of $\anumu$-CC($\pi$).   The value for $\phi^{\anu}$ indicates that $A_{3}$ and $A_{1}$ are, on average,
roughly $90^{\circ}$ out of phase.   These observations are consistent with a resonant $I=3/2$ amplitude
whose phase is rotating counterclockwise through $\pi/2$ (at the $\Delta$ peak), while the phase of the
nonresonant $I=1/2$ amplitude remains stationary near $0^{\circ}$.

 \subsection{Bubble chamber measurements}
 \label{subsec:BC-measurements}

The isospin decomposition reported here was originally utilized by bubble
chamber experiments of the 1970s and 1980s.    A full determination of $R^{\anu}$ and $\phi^{\anu}$ 
 for the $\anumu$-CC($\pi$) channels was carried out using the Gargamelle bubble chamber filled with 
 a light propane-freon mixture~\cite{Gargamelle-1979}.    Table~\ref{tab:Anu-ratio-phase} compares the present MINERvA measurement
 with the Gargamelle result.   The measurement precisions are seen to be roughly comparable, reflecting the fact that
 MINERvA's statistical advantage (factor $\sim$2.2 in event candidates) is partially offset by systematic uncertainties that
 are larger than those incurred with the bubble chamber technique.   Together,
 the two experiments give a very consistent picture of the isospin composition of  
 $\anumu$-CC($\pi$) channels.

\begin{table}
\begin{center}
\begin{tabular}{cccccccc}
\hline
\hline
\rule{0pt}{2.5ex} Experiment  &$\anu$ flux  &$W$ & $R^{\anu}$  & $\phi^{\anu}$  \\
\rule{0pt}{2.2ex} medium& \,[GeV] & [GeV] &     & degrees  \\ 
\cline{1-5} 
&&&&\\ [-7pt]
Gargamelle\cite{Gargamelle-1979} & $\sim\,$0.5 - 10.0& $\le$ 1.8  &~1.14$\pm 0.23$ & 94$\pm 13^{\circ}$  \\ [4pt]
propane-freon & peak: 1.5 & $\le$ 1.4 &~\,0.98$\pm 0.20$& \{$90^{\circ}$\}  \\ [3pt]
&&&&\\ [-7pt]
&&&&\\ [-7pt]
MINERvA & $\sim\,1.5 - 10.0$& $\le$ 1.8  &~0.99$\pm 0.19$ & 93$\pm 7^{\circ}$  \\ [4pt]
hydrocarbon & peak: 3.0 &   &       &   \\ [3pt]
\hline
\hline
\end{tabular}
\caption{Antineutrino measurements of relative strength, $R^{\anu}$, and relative phase, 
$\phi^{\anu}$, for isospin 1/2 and 3/2 
amplitudes of $\anumu$-CC($\pi$) production.   Results of this work (lower rows, leftmost columns) are
in good agreement with values obtained four decades ago using the Gargamelle bubble chamber.}
\label{tab:Anu-ratio-phase}
\end{center}
\end{table}

Under the assumption that the $\Delta S = 0$ charged current operator is charge symmetric, antineutrino reactions 
$\anu + (-I^{i}_{3}) \rightarrow \mu^{+} + (-I^{f}_{3})$ may be related to neutrino reactions 
$\nu + (I^{i}_{3}) \rightarrow \mu^{-} + (I^{f}_{3})$, where the initial and final hadronic systems are labeled by their $I_{3}$ values.
This relation motivates a comparison of the isospin amplitude relations of the present work to those obtained 
by the large bubble chamber experiments in analysis of neutrino-induced single pion production.    
Decomposition of the three exclusive channels of $\numu$-CC($\pi$) proceeds as previously described, 
but with cross sections \eqref{xsec-exclusive-channel-1}, \eqref{xsec-exclusive-channel-2}, and \eqref{xsec-exclusive-channel-3} 
replaced by $\sigma(\mu^- \pi^+ p)$, $\sigma(\mu^- \pi^+ n)$, 
and $\sigma(\mu^- \pi^0 p)$ respectively.   The bubble-chamber measurements 
for $R^{\nu}$ and $\phi^{\nu}$ of neutrino-induced $\pi N$ systems are summarized in Table~\ref{tab:Nu-ratio-phase}. 
As with the $\anumu$-CC($\pi$) results, the $\numu$ measurements also find the $I=1/2$ amplitude to be sizable
relative to the resonant $I=3/2$ amplitude, and indicate the two amplitudes to be $90^{\circ}$ out of phase on average.

\begin{table}
\begin{center}
\begin{tabular}{cccccccc}
\hline
\hline
\rule{0pt}{2.5ex} Experiment  &$\nu$ flux  &$W$ & $R^{\nu}$  & $\phi^{\nu}$  \\
\rule{0pt}{2.2ex} medium& \,[GeV] & [GeV] &     & degrees  \\ 
\cline{1-5} 
&&&&\\ [-7pt]
Gargamelle\cite{Pohl-nu-cc-pi-1979}& $\sim\,$0.5 - 10.0& $\le$ 1.4  & 0.71$\pm 0.14$ & 75$^{+12^{\circ}}_{-16^{\circ}}$  \\ [4pt]
propane-freon& peak: 1.5 &all data & 1.03$\pm 0.15$& 73$^{+12^{\circ}}_{-10^{\circ}}$ \\ [3pt]
&&&&\\ [-7pt]
&&&&\\ [-7pt]
BNL 7' BC\cite{Baker-nu-cc-pi-1981}& $< 3.0$ &$\le$ 1.4  & 0.60$\pm 0.07$ & 90$\pm 11^{\circ}$  \\ [4pt]
deuterium& peak: 1.0 &$\le$ 1.6 & 0.79$\pm 0.05$& 95$\pm 7^{\circ}$  \\ [3pt]
                 &  &all data  & 0.89$\pm 0.05$& 97$\pm 6^{\circ}$  \\ [3pt]
&&&&\\ [-7pt]
&&&&\\ [-7pt]
ANL 12' BC\cite{Radecky-1982}&$< 1.5$ & $\le$ 1.4  & 0.68$\pm 0.04$ & 90.7$\pm 4.6^{\circ}$  \\ [4pt]
deuterium& peak: 0.5 & $\le$1.6 & 0.75$\pm 0.04$& 92.0$\pm 4.1^{\circ}$  \\ [3pt]
\hline
\hline
\end{tabular}
\caption{Neutrino bubble chamber measurements of relative strength and phase for the isospin 1/2 and 3/2 
amplitudes of neutrino-induced $CC(\pi$) production.   Values obtained for neutrino-induced
$R^{\nu}$ and $\phi^{\nu}$ are similar to those reported in Table~\ref{tab:Anu-ratio-phase} for antineutrino single-pion production. }
\label{tab:Nu-ratio-phase}
\end{center}
\end{table}

\begin{figure}
  \begin{center}
      \includegraphics[width=9.0cm]{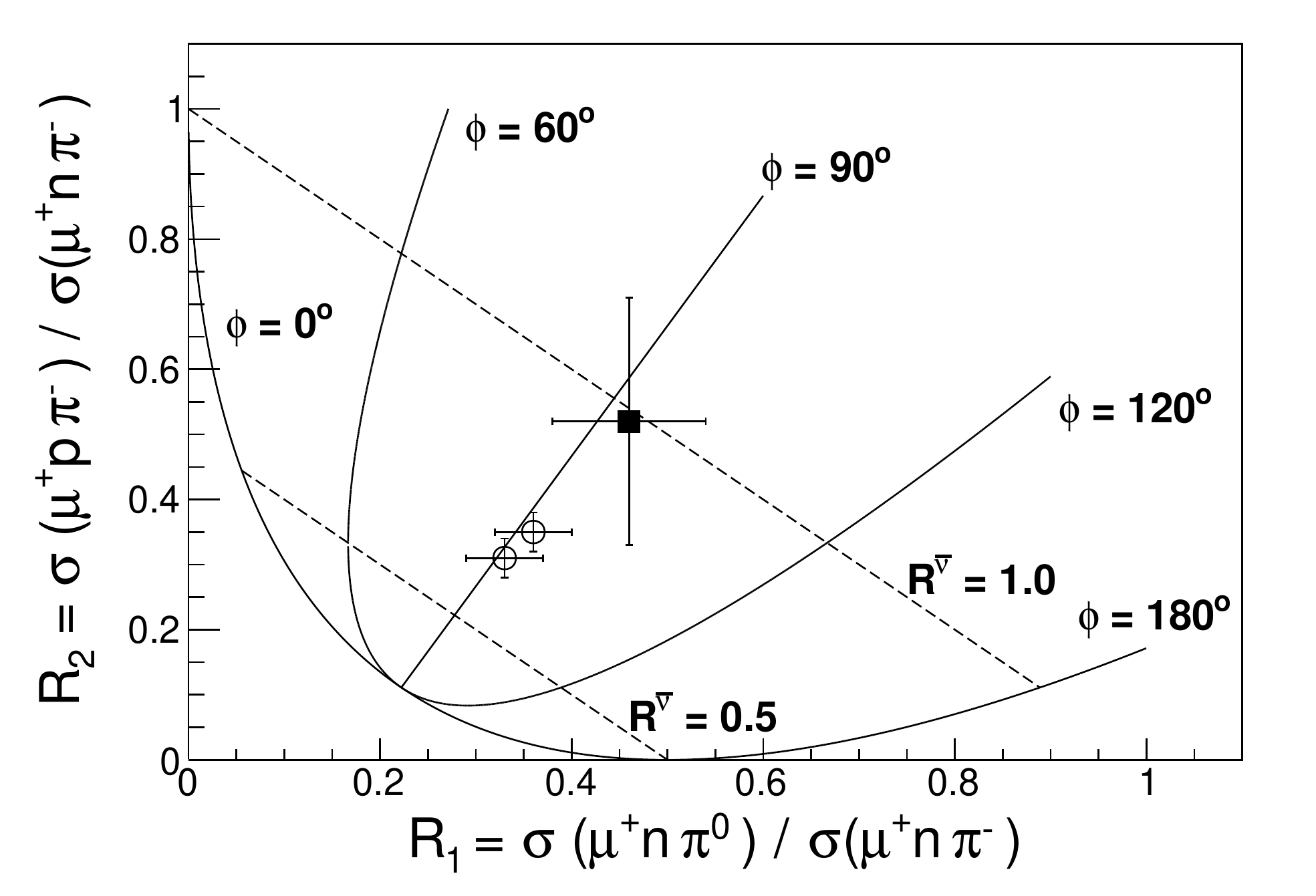}
 \caption{Plot of the cross-section ratios $R_{2}$ versus $R_{1}$ for selected $\anumu$ and $\numu$ data.   Dashed lines denote
constant values of $|A_{1}| / |A_{3}|$ and solid-line curves denote values of the relative phase.  
The MINERvA measurements (solid square), as with Gargamelle (not plotted; see Table \ref{tab:Anu-ratio-phase}) 
show that $A_{1}$ and $A_{3}$, averaged over a wide-band $\anu$ flux, are of similar strength and devoid of interference. 
Results obtained with $\numu$-CC($\pi$) reactions (open circles, from overlapping samples) indicate $|A_{3}| > |A_{1}|$ in neutrino
samples at lower incident energies and lesser reach in $W$~\cite{Radecky-1982}.}
    \label{Fig17}
  \end{center}
\end{figure}

A discernible trend in the neutrino results is that higher reach in $W$ correlates with larger $R^{\nu}$ values.
This is understandable because, above $W > 1.4$\,GeV, the $\Delta(1232)$ contribution is diminished while $I = 1/2$ baryon resonances
gain strength.    The MINERvA data contain a relatively large contribution from events with $W$ between 1.4 - 1.8 GeV compared
to the ANL and BNL data sets, and this may be the reason why $R^{\anu}$ of MINERvA is larger than $R^{\nu}$ as measured
by ANL and BNL.    

A convenient way to compare measurements of the relative magnitude and phase of $A_{1}$ versus $A_{3}$
is with the diplot shown in Fig.~\ref{Fig17}.    The plot maps measurements of the cross-section ratios $R_{1}$ and $R_{2}$
onto a coordinate grid of slanted dashed lines and solid-line curves that denote values of $R^{\anu}$ and $\phi^{\anu}$ respectively.
The MINERvA and Gargamelle antineutrino measurements lie within $1 \sigma$ of ($R^{\anu}$, $\phi^{\nu}$) $\simeq (1.0, 90^{\circ})$,
indicating the amplitude strengths to be nearly equal and non-interfering ($\cos\phi^{\anu} \simeq 0$).
The neutrino measurements, working with lower-$W$ samples, also lie along the $\phi = 90^{\circ}$ axis but at $R^{\nu}$ 
values distinctly less than 1.0.   The plot suggests that the representation point for a CC($\pi$) sample migrates upward along
$\phi^{\nu} = 90^{\circ})$, as the average $W$ of the sample is increased.

\section{Conclusions}
\label{sec:Conclude}

A study of semi-exclusive $\anumu$-CC($\pi^{-}$) scattering on hydrocarbon is reported using $\anumu$ interactions with $E_{\anu}$
ranging from $\sim\,1.5$ to 10 GeV, with final-state $W < 1.8$\,GeV.
This is the first experiment working in the few-GeV region of incident $\anumu$ to report differential cross sections for $\mu^{+}$ 
and $\pi^{-}$ kinematic variables $\theta_{\mu}$, $p_{\mu}$, $T_{\pi}$,
and $\theta_{\pi}$, while also reporting cross sections as functions of $E_{\anu}$ and $Q^{2}$.
Data summary tables for these measurements that may facilitate phenomenological investigations
are available in the Supplement~\cite{Supplement}.    

Measured differential cross sections are compared to predictions 
based upon the GENIE, NuWro, and GiBUU-2017 event generators.  The predictions generally reproduce 
the shapes of the differential cross sections, with $d\sigma/d\theta_{\pi^{-}}$ being the sole exception.
 The event generators differ with respect to predictions for absolute event rate.   The GENIE-based simulation
gives the highest event rate and its prediction exceeds the observed data rate by 8\%.   

The shape of the pion $T_{\pi}$ differential cross section is considered in light of
GENIE's effective cascade treatment of processes that comprise pion FSI.   
The modeling provides a detailed picture for the $d\sigma/T_{\pi}$ distribution that is consistent with
the data (Fig.~\ref{Fig12}).   This same picture suggests that adjustments to pion FSI elastic and inelastic
scattering that promote emission into smaller, more forward angles may be in order (Fig.~\ref{Fig13}).
For $d\sigma/dQ^{2}$, neither the data nor the generator curves exhibit a turn-over in the distribution at very-low $Q^2$.
This observation contrasts with distribution turn-over for $Q^2 < 0.20$\,GeV$^2$ that occurs in MINERvA measurements
for $\anumu$-CC($\piz)$~\cite{Trung-pion} and $\numu$-CC($\piz)$ channels~\cite{Carrie-pion, Altinok-2017}.

The signal sample has been decomposed into $\anumu$ interactions 
of four kinds, with exclusive reactions \eqref{exclusive-channel-1} and \eqref{exclusive-channel-2} 
being the major contributors.  Flux-averaged quasi-free nucleon
scattering cross sections are presented in  Eqs.~\eqref{xsec-exclusive-channel-1} and \eqref{xsec-exclusive-channel-2}.
The flux-averaged cross section \eqref{xsec-exclusive-channel-3} is 
extracted from the published MINERvA measurement of $\anumu$-CC($\pi^{0}$).
These three $\anumu$-nucleon cross sections are used to carry out an isospin decomposition of
 CC single pion production initiated by anti-neutrino (non-coherent) interactions.  
The relative magnitude and phase of isospin amplitudes $A_{1}$ and $A_{3}$ presented in Eq.~\eqref{magnitude-phase-1}
are in agreement with the pioneering Gargamelle measurement~\cite{Gargamelle-1979}. 

In summary, the measurements of this work 
introduce a wealth of new information about $\anumu$-CC($\pi)$, an antineutrino interaction channel that features prominently 
in data samples being recorded by the long-baseline experiments.   These results pave the way for 
more precise determinations of the fundamental parameters that govern flavor oscillations of neutrinos and antineutrinos.

\section*{Acknowledgments}

This document was prepared by members of the MINERvA Collaboration using the resources of the Fermi National Accelerator Laboratory (Fermilab), a U.S. Department of Energy, Office of Science, HEP User Facility. Fermilab is managed by Fermi Research Alliance, LLC (FRA), acting under Contract No. DE-AC02-07CH11359.
These resources included support for the MINERvA construction project, and support
for construction also
was granted by the United States National Science Foundation under
Award No. PHY-0619727 and by the University of Rochester. Support for
participating scientists was provided by NSF and DOE (USA); by CAPES
and CNPq (Brazil); by CoNaCyT (Mexico); by Proyecto Basal FB 0821, CONICYT PIA ACT1413, Fondecyt 3170845 and 11130133 (Chile); 
by CONCYTEC (Consejo Nacional de Ciencia, Tecnología e Innovación Tecnológica), DGI-PUCP (Dirección de Gestión de la Investigación  - Pontificia Universidad Católica del Peru), and VRI-UNI (Vice-Rectorate for Research of National University of Engineering) (Peru);
and by the Latin American Center for Physics (CLAF); NCN Opus Grant No. 2016/21/B/ST2/01092 (Poland); by Science and Technology Facilities Council (UK).  We thank the MINOS Collaboration for use of its near detector data. Finally, we thank the staff of
Fermilab for support of the beam line, the detector, and computing infrastructure.

This manuscript has been authored by Fermi Research Alliance, LLC under Contract No. DE-AC02-07CH11359 with the U.S. Department
of Energy, Office of Science, Office of High Energy Physics. The United States Government retains and the publisher, by
accepting the article for publication, acknowledges that the United States Government retains a non-exclusive, paid-up,
irrevocable, world-wide license to publish or reproduce the published form of this manuscript, or allow others to do so, for
United States Government purposes.


\end{document}